\documentclass[journal]{IEEEtran}
\usepackage{amsmath,amsfonts,amsthm,amssymb,bm}
\usepackage{algorithmic}
\usepackage{array}
\usepackage[caption=false,font=normalsize,labelfont=sf,textfont=sf]{subfig}
\usepackage{textcomp}
\usepackage{stfloats}
\usepackage{url}
\usepackage{verbatim}
\usepackage{graphicx}
\usepackage{bbding}
\usepackage{multirow}
\usepackage{cite, bbm, xcolor}
\usepackage{refcount}
\usepackage{flafter}
\usepackage{makecell}
\usepackage{diagbox}
\usepackage{booktabs}
\usepackage{algorithm}

\newtheorem{proposition}{Proposition}
\newtheorem{lemma}{Lemma}

\theoremstyle{definition}

\usepackage{subcaption}
\usepackage{float}
\def\BibTeX{{\rm B\kern-.05em{\sc i\kern-.025em b}\kern-.08em
    T\kern-.1667em\lower.7ex\hbox{E}\kern-.125emX}}
\usepackage{balance}
\begin{document}
\title{LLM Enabled Beam Training for Pinching Antenna Systems (PASS)}

\author{
        Deqiao Gan, Xiaoxia Xu, Xiaohu Ge, \textit{Senior Member, IEEE}, and Yuanwei Liu, \textit{Fellow, IEEE}
        \thanks{D. Gan and X. Ge (Corresponding author) are with the School of Electronic Information and Communications, Huazhong University of Science and Technology, Wuhan 430074, Hubei, China. (e-mail: gandeqiao@hust.edu.cn, xhge@mail.hust.edu.cn).}
        \thanks{X. Xu is with the School of Electronic Engineering and Computer Science, Queen Mary University of London, London E1 4NS, U.K. (email: x.xiaoxia@qmul.ac.uk).}
        \thanks{Y. Liu is with the Department of Electrical and Electronic Engineering, The University of Hong Kong, Hong Kong (e-mail: yuanwei@hku.hk).}
}


\maketitle

\begin{abstract}
    To enable intelligent beam training, a large language model (LLM)-enabled beam training framework is proposed for the pinching antenna system (PASS) in downlink multi-user multiple-input multiple-output (MIMO) communications.
    A novel LLM-based beam training supervised learning mechanism is developed, allowing context-aware and environment-adaptive probing for PASS to reduce overheads. Both single-user and multi-user cases are considered. 1) For single-user case, the LLM-based pinching beamforming codebook generation problem is formulated to maximize the beamforming gain. Then, the optimal transmit beamforming is obtained by maximum ratio transmission (MRT). 2) For multi-user case, a joint codebook generation and beam selection problem is formulated based on the system sum rate under the minimum mean square error (MMSE) transmit beamforming. The training labels for pinching beamforming are constructed by selecting the beam combination that maximizes system performance from each user's Top-$S$ candidate beams. Based on pretrained Generative Pre-trained Transformers (GPTs), the LLM is trained in an end-to-end fashion to minimize the cross-entropy loss. Simulation results demonstrate that: i) For single-user case, the proposed LLM-enabled PASS attains over 95$\%$ Top-1 accuracy in beam selection and achieves 51.92$\%$ improvements in beamforming gains compared to conventional method. ii) For multi-user case, the proposed LLM-enabled PASS framework significantly outperforms both the LLM-based massive MIMO and conventional PASS beam training, achieving up to 57.14$\%$ and 33.33$\%$ improvements in sum rate, respectively.
\end{abstract}


\section{Introduction}
\IEEEPARstart{T}{o} accommodate the ultra-high-speed and massive connectivity requirements for next-generation wireless networks, millimeter-wave (mmWave)/terahertz (THz) communication has emerged to leverage vast spectrum resources in high-frequency bands \cite{yang2025pinching, kit2021radio}.
However, the high-frequency propagation inherently results in high propagation loss, posing a critical challenge for reliable communication. To overcome this limitation, massive multiple-input multiple-output (MIMO) antenna arrays have been widely adopted, employing highly directional beamforming techniques to compensate for severe path loss \cite{heath2018foundations,2013MIMO}.
Although conventional massive MIMO systems have considerable spatial multiplexing gains, their reliance on rigid large-scale antenna arrays and static beamforming architectures limit their ability to effectively respond to user mobility and dynamic line-of-sight (LoS) blockage. To address these constraints, several flexible-antenna techniques have been developed, including reconfigurable intelligent surfaces (RISs) \cite{yuanwei2021RIS,tang2021RIS}, fluid antennas \cite{kit2021fluid,new2024fluid} and movable antennas \cite{zhu2024movable,zhu2025movable}.

Conceptually, pinching antenna system (PASS) serves as a new form of flexible-antenna technology that offers greater flexibility and scalability. PASS was first prototyped by NTT DOCOMO \cite{2022NTTDOCOMO}, and then introduced into wireless communications \cite{ding2024pass}.
An earlier concept known as surface-wave communication superhighways was proposed in \cite{kit2021radio}, which leverages the similar principles of in-waveguide propagation on reconfigurable surfaces.
Different from conventional reconfigurable-antenna architectures, PASS leverages dielectric and extensible waveguides embedded with separately configurable PAs, which can be dynamically activated and repositioned along the waveguide \cite{yuanwei2025pass, gan2025NOMAPASS}. 
With reconfigurable waveguides and pinching antennas (PAs) for in-waveguide propagation, PASS can reduce path loss and improve transmission efficiency \cite{liu2024path, chu2024propagation}.
Since PASS enables the flexible deployment of PAs close to end users, it establishes robust LoS connections. Moreover, this spatial adaptability allows PASS to effectively mitigate large-scale path loss and overcome coverage gaps \cite{wang2025pass,xu2025pass2}. Hence, PASS enables a novel form of beamforming, referred to as pinching beamforming, where both the large-scale path loss and the phase of the transmitted signal can be continuously reconfigured in the spatial domain \cite{ouyang2025pass, xu2025antenna}. Distinctly, PASS demonstrates several unique advantages:
\textit{1) Adjustable position and beamforming}: PAs can be readily repositioned along the dielectric waveguide to accommodate real-time spatial and service demands, enabling unprecedented adaptability over large physical areas. By activating PAs at the desired positions, PASS can achieve adjustable beamforming.
\textit{2) Path loss configurations and LoS support}: The adjustment of PAs allows PASS to dynamically control large-scale path loss while maintaining strong LoS links, significantly reducing the risk of high-frequency link blockage \cite{ding2025losblockage}.

The strong directionality of narrow beams formed by extensive arrays leads to increased sensitivity to user mobility and environmental obstructions, necessitating precise alignment between transmitted and received beams. The dynamic PA location require tracking and adapting to user movements, which results in continuously varying channel environments, complicating both channel estimation and prediction. Furthermore, the unpredictability of user and PA positions makes it difficult to leverage historical data for accurate channel modeling, especially since the beam direction and PA locations are tightly coupled and directly impact system performance. Despite the significant advantages of PASS in terms of spatial adaptability and flexible beamforming, several challenges remain as follows:
\begin{itemize}
    \item \emph{Large Beam Training Overheads}: To achieve high selection accuracy, an exhaustive search over a large beamspace is often required, which results in prohibitively high computational complexity and substantial beam training overhead.
    \item \emph{User Mobility}: The dynamic nature of user locations necessitates frequent and rapid adjustment of PA positions to continuously track and serve mobile users, thereby increasing the complexity of system control and adaptation.
    \item \emph{Limited Accuracy}: It remains an open problem how to fully exploit the integrated information from multiple PAs and auxiliary environmental sensing from multimodal sensors, e.g., camera images, light detection and ranging (LiDAR), Global Positioning System (GPS), to further enhance the accuracy and reliability of beam training in practical deployments.
\end{itemize}

To address these challenges, the development of beam training for PASS framework is crucial. The authors of \cite{lv2025beam} proposed a three-stage beam training scheme for PASS, which utilized a coarse-fine-search strategy for beam training. Traditional hierarchical beam training schemes \cite{qi2020hierarchical, wu2023beam}, originally designed for static arrays, struggle to accommodate the added spatial degrees of freedom in PASS. Moreover, most near-field beam training techniques, whether implemented at the base station (BS) or via RIS, assume fixed antenna placements, making them unsuitable for the highly flexible and reconfigurable PASS paradigm.

Large language models (LLMs) play a critical role in beam training \cite{shao2024wirelessllm,jiang2024llm}. On the one hand, LLMs have strong representation learning capabilities, and can automatically extract and fuse complex spatial, context, and multimodal features to provide deep information support for accurate beam training. On the other hand, LLM can generalize to unknown environments and dynamic user distribution, realize adaptive reasoning for beam selection, and effectively deal with complex and nonlinear relationships that are difficult to tackle in traditional models. The massive MIMO beam training scheme that utilizes LLMs to fuse multimodal and channel information for mmWave beam selection was proposed by \cite{zheng2025beamllm}. The authors of \cite{zhou2024LLM} designed the application of LLM to wireless communications, enabling unified multi-task optimization and cross-modal reasoning for beam management.
However, there is little literature researching LLM-driven beam training for PASS. This context motivates the development of new channel acquisition and beam training approaches tailored to the spatial dynamics of PASS, utilizing both physical models and LLM to optimize system performance.


In this paper, we propose a novel beam training PASS framework enabled by LLM in massive MIMO scenarios. Specifically, we develop the end-to-end multi-modal sensing empowered beam training framework based on LLM for PASS, which integrates the multi-modal sensing information, including camera images, LiDAR, and GPS, to generate high-dimensional tokens by for comprehensive perception of radio environment. Leveraging the generalization and representation capabilities of LLM, we design an LLM-based codebook generation and beam selection mechanism for beam training of PASS. We consider single-user and multi-user cases. For single-user case, we maximize the beamforming gain by training LLM to generate pinching beamforming codebook for optimal beam configuration. Then, the transmit beamforming for communication is obtained by maximum ratio transmission (MRT). For multi-user case, we maximize the system sum rate by jointly optimizing pinching beamforming codebook generation and beam selection under the minimum mean square error (MMSE) transmit beamforming strategy. We propose a supervised end-to-end training mechanism to tailor the LLM. The key contributions of this work are summarized as follows:
\begin{enumerate}
    \item We propose a novel beam training framework for PASS, empowered by multimodal sensing information, including camera image, LiDAR, and GPS data. We develop an LLM-based supervised learning mechanism to jointly perform codebook generation and beam selection. The multimodal features are integrated into high-dimensional tokens to achieve comprehensive environmental understanding. The pinching beamforming codebook, which samples a set of pinching antenna positions for sequential probing, is then generated based on the high-dimensional tokens for efficient beam training. We consider both single-user and multi-user cases.
    \item For single-user case, we formulate an LLM-based pinching beamforming codebook generation problem to maximize the beamforming gain. The optimal pinching beamforming is selected from the codebook generated by LLM. Based on the selected pinching beamforming, the optimal transmit beamforming is obtained by MRT. 
    \item For multi-user case, we formulate a joint pinching beamforming codebook generation and beam selection problem to maximize the system sum rate, which can enhance beamforming gain and mitigate interference by combining MMSE transmit beamforming strategy. From Top-$S$ pinching beamforming candidates of each user, we search the optimal combination for multi-user communication as training labels for efficient training. Built on pretrained Generative Pre-Trained Transformer (GPT) modules, the LLM is further trained through an end-to-end supervised training process by minimizing cross-entropy loss.
    \item Numerical simulation results verify the effectiveness of the proposed LLM-enabled PASS beam training, which demonstrates that: i) In single-user case, the proposed LLM-enabled PASS achieves 51.92$\%$ improvements in beamforming gain compared to standard PASS, and achieves over 95$\%$ in Top-1 accuracy of beam selection. ii) In multi-user case, the proposed LLM-enabled PASS increases the system sum rate by 57.14$\%$ and 33.33$\%$ compared to LLM-enabled massive MIMO and the conventional PASS beam training method. 
\end{enumerate}



The remainder of this paper is organized as follows: Section II introduces the system model of LLM-enabled PASS framework and the problem formulation. Section III and Section IV focus on the beam training framework based on LLM for PASS, consists of multimodal feature extraction, codebook generation and beam selection, for single-user and multi-user cases, respectively. Numerical results are analyzed in Section V. The conclusion is shown in Section VI.

\textit{Notations}: $\mathbb{C}^{M \times 1}$ and $\mathbb{R}^{M\times 1}$ denote the space of $M \times 1$ complex valued vectors and real valued vectors, respectively. 
$\text{blkdiag}(\mathbf{x})$ denotes a block-diagonal matrix. The variable, vector, and matrix are denoted by $x$, $\mathbf{x}$, and $\mathbf{X}$, respectively.
$\text{Re}\left\{x\right\}$ and $\text{Im}\left\{x\right\}$ denote the real and image parts of $x$, and $x^{H}$ is the complex conjugate number of $x$. 
$\mathbf{X}^{T}$ and $\mathbf{X}^{H}$ denote the transpose and the Hermitian matrix. 
$\mathcal{CN}(0,\sigma^2)$ represents the distribution of a circularly symmetric complex Gaussian variable (CSCG) with zero mean and $\sigma^2$ variance.
$\mathcal{TOP}_S$ represents the true label included among the top-$S$ predicted beamforming vectors with the $S$ highest values. $\mathbf{I}_k$ represents the k-order identity matrix.

\section{System Model}

\begin{figure}[!t]
    \centering
    \includegraphics[width=3.6in]{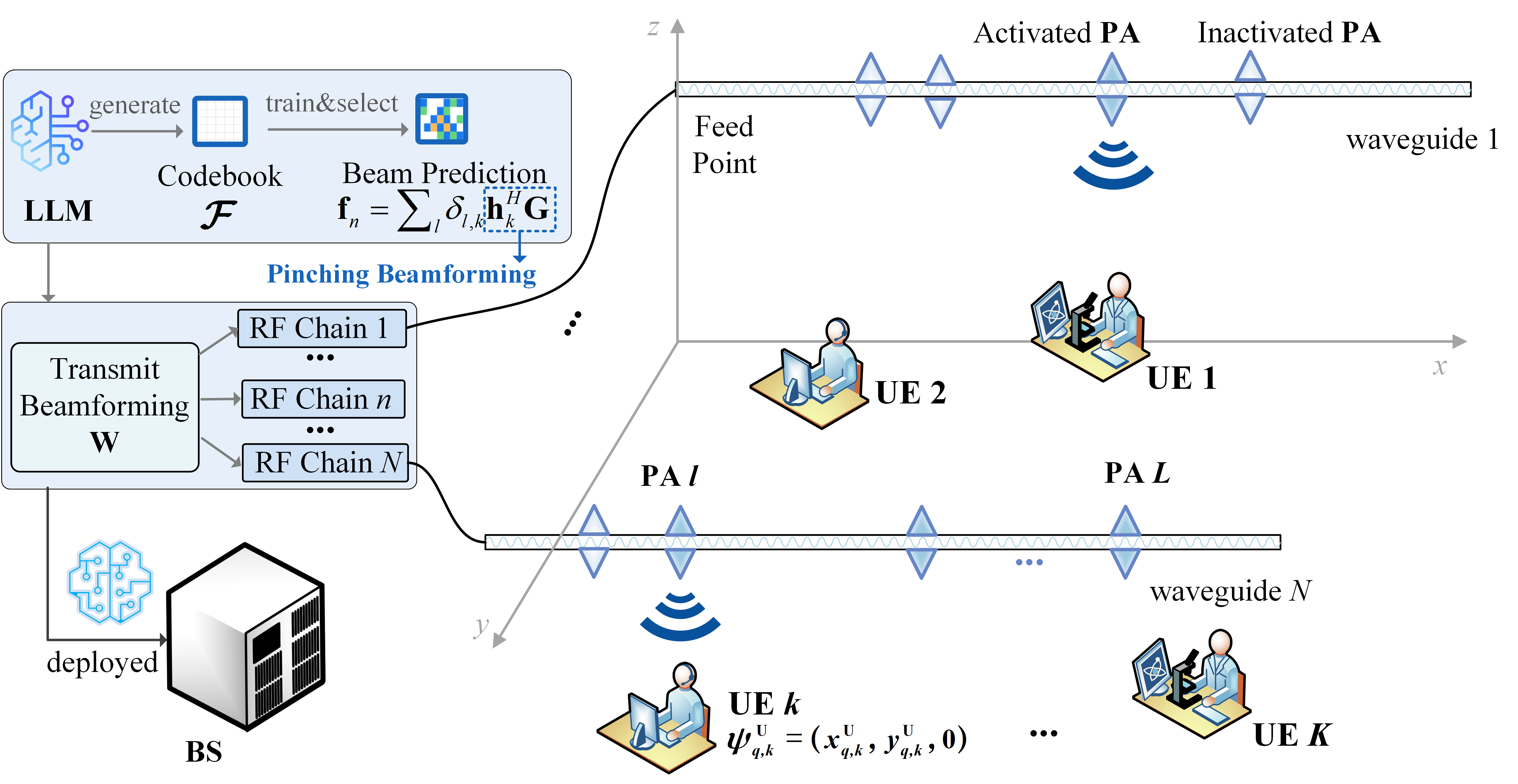}
    \caption{System model: The proposed LLM-enabled PASS framework.}
    \label{systemmodel}
\end{figure}
Consider an LLM-enabled PASS framework in MIMO communication scenario, where one BS serves a set of $\mathcal{K}=\{1,2,\dots,K\}$ of users, $k \in \mathcal{K}$. This framework comprises dielectric waveguides $n \in \mathcal{N}=\{1,2,\dots,N\}$ each pinched along $L$ pinching antennas (PAs), indexed by $l \in \mathcal{L} = \{1,2,\dots,L\}$. The proposed LLM-enabled PASS framework for beam training is shown as Fig. \ref{systemmodel}. The waveguide is connected to a dedicated radio frequency (RF) chain, which converts the signal multiplexed at the baseband and feeds it into the waveguide. And the total number of PAs is $M = N \times L$. Denote that both waveguides and PAs are installed at a fixed height of $\text{d}_0$, users are initialized at the height of 0. The length of each waveguide is denoted by $x^{\max}$, and the distance between two adjacent PAs is defined as $y_n=(n-1)y_0$. Let the three-dimensional Cartesian coordinates of feed point for waveguide $n$ as ${\bm{\psi}}^{\text{W}}_n=[0,y_n,\text{d}_0]$. The PASS spans across the rectangular area with a size of $(x^{\max} \times y^{\max})$, which is the same as the distribution coverage of users. Thus, the location of user $k$ is denoted by ${\bm{\psi}}^{\text{U}}_k=[x_k,y_k,0], \forall k \in \mathcal{K}$, where $0 \le x_k \le x^{\max}$ and $0 \le y_k \le y^{\max}$. The location of flexible PA $l$ on waveguide $n$ along $x$-axis is defined by $x_{n,l}$. Let $\mathbf{x}_{n}=[x_{n,1},x_{n,2},...,x_{n,L}]^T \in \mathbb{R}^{L \times 1}$ represent the $x$-axis locations of PAs on the waveguide $n$, and $\mathbf{X}=[\mathbf{x}_{1},\mathbf{x}_{2},...,\mathbf{x}_{N}] \in \mathbb{R}^{L\times N}$ typify the total PA location matrix along $x$-axis, where $\forall n \in \mathcal{N}, \forall l \in \mathcal{L}$. 

\subsection{PASS Channel Model}
The effective response vector from waveguide $n$ to PAs can be written as
\begin{equation}
    \label{responsevector}
    g_{n,l}(x_{n,l})=\frac{1}{\sqrt{L}}e^{-i\frac{2\pi}{{\lambda}_g}\left|{\bm{\psi}}_{n,0}^{\text{PA}}-{\bm{\psi}}_{n,l}^{\text{PA}}\right|}=\frac{1}{\sqrt{L}}e^{-i\frac{2\pi}{{\lambda}_g}x_{n,l}},
\end{equation}
where $\lambda_g = \frac{\lambda}{n_{\text{eff}}}$ represents the guided wavelength, $\lambda$ is the wavelength, $n_{\text{eff}}$ represents the effective refractive index of the dielectric waveguide, and $\frac{1}{\sqrt{L}}$ means the amplitude of the transmitted signal, that is, the power allocation coefficient for $L$ PAs along waveguide $n$. Additionally, $|{\bm{\psi}}_{n,0}^{\text{PA}}-{\bm{\psi}}_{n,l}^{\text{PA}}| = x_{n,l}$ denotes the distance from the feed point of waveguide $n$ to PA $l$. Then, the overall in-waveguide channel matrix can be expressed as
\begin{equation}
    \label{blkdiag}
    \begin{aligned}
        \mathbf{G}(\mathbf{X}) \left.=\left[
        \begin{array}{cccc}
            \mathbf{g}_1\left(\mathbf{x}_1\right) & \mathbf{0} & \ldots & \mathbf{0} \\
            \mathbf{0} & \mathbf{g}_2\left(\mathbf{x}_2\right) & \ldots & \mathbf{0} \\
                \vdots & \vdots & \ddots & \vdots \\
            \mathbf{0} & \mathbf{0} & \ldots & \mathbf{g}_N\left(\mathbf{x}_N\right)
        \end{array}
        \right.\right].
    \end{aligned}
\end{equation}

Consequently, the geometric free-space spherical propagation model is employed to characterize the channel vector between PA $l$ on waveguide $n$ and user $k$, which is expressed as
\begin{equation}
    \label{channelvector}
    h^H_{n,l,k} (x_{n,l}) = \frac{\eta e^{-i{\kappa}\left|{\bm{\psi}}_k^{\text{U}}-{\bm{\psi}}_{n,l}^{\text{PA}}\right|}}{\left|{\bm{\psi}}_k^{\text{U}}-{\bm{\psi}}_{n,l}^{\text{PA}}\right|}, 
\end{equation}
where $\kappa=\frac{2\pi}{\lambda}$ is the wave-domain number. Hence, the channel vector from the PAs along wavegudie $n$ is denoted by $\mathbf{h}^H_{n,k} (\mathbf{x}_n) \in \mathbb{C}^{1 \times L}$, and the channel matrix for all PAs is denoted by $\mathbf{h}^H_{k}(\mathbf{X}) = [\mathbf{h}^H_{n,k} (\mathbf{x}_1),\mathbf{h}^H_{n,k} (\mathbf{x}_2),\dots,\mathbf{h}^H_{n,k} (\mathbf{x}_N)] \in \mathbb{C}^{1 \times M}$.
To reflect practical propagation conditions, this paper considers both the line-of-sight (LoS) and the non-line-of-sight (NLoS) \cite{ding2025losblockage}. Since the PA and UE are movable and the distance between them might be long-range, thus, we build a blockage model for PASS to contemplate the multi-user interference link that may be blocked. We consider two cases, i.e., single-user and multi-user cases, for this model. The indicator function of LoS blockage $\delta_{l,k}$ is introduced into this model. If blockage exists between user $k$ and PA $l$, $\delta_{l,k} = 0$. In contrast, $\delta_{l,k} = 1$, which can be expressed as
\begin{equation}
    \label{blockage}
    \delta_{l,k}=
    \begin{cases}
       1, & \text{LoS} \\
       0, & \text{NLoS} 
\end{cases},
\end{equation}
for user $k$, the channel vector with LoS blockage can be formulated as $\widetilde{h}^H_{n,l,k} (x_{n,l})=\delta_{l,k} h^H_{n,l,k} (x_{n,l})$.

\subsection{PASS Signal Model}

\subsubsection{Channel Probing}
Pinching beamforming enables flexible spatial configuration of antenna arrays, allowing dynamic adaptation to user locations and channel conditions, thereby significantly enhancing beamforming gain and overall system performance in PASS.
The pinching beamforming codebook is constructed on the BS side, including all feasible PA activation positions called as codewords. Each $\mathbf{h}^H_k\mathbf{G}$ represents the activating schedule of PAs. With LoS blockage, the physical gain of pinching beamforming along waveguide $n$ under each codeword needs to be multiplied by the indicator factor, which is denoted by
\begin{equation}
    \label{pinchingbeamforming}
    \mathbf{f}_n(\mathbf{X})= \sum_l\delta_{l,k}\mathbf{h}^H_k(\mathbf{X})\mathbf{G}(\mathbf{X}),
\end{equation}
where the predefined pinching beamforming codebook $\mathbf{F}=\{\mathbf{f}_1,..., \mathbf{f}_N\}$ and $\mathbf{f}_n \in \mathbb{C}^{1\times N}$. If the main components of the beam direction are blocked (such as the corresponding $\delta_{l,k}=0$), the actual gain of this codeword is reduced to 0. At the time slot $t \in \mathcal{T}=\{1,2,\dots,T\}$, the pinching beamforming vector $\mathbf{f}_n[t]$ is determined by BS according to the information of the codebook during 0-$(t-1)$ time slots.
The PASS utilizes multiple dielectric distributed waveguides with flexible PAs arranged on each waveguide. Each beam is no longer simply equivalent to an angle of departure (AoD), but is determined by the spatial location of PAs with an activation scheme.

During the channel probing phase, the system sequentially transmits probing signals using each pinching beamforming configuration from the above generated codebook. At this stage, the transmit digital beamforming matrix is set as the identity matrix $\mathbf{W}^{\text{prob}}$. At channel probing phase, transmit signal can be expressed as
\begin{equation}
    \label{probingsignal}
    s^{\text{prob}}_{k}[t]= \sqrt{\alpha_{k}}\mathbf{W}^{\text{prob}}\mathbf{s}^{\prime}_{n,k}[t], \forall t \in \mathcal{T},
\end{equation}
where $\alpha_{k}$ is the power allocation coefficient for UE $k$, $\mathbf{s}^{\prime}_{n,k}[t]$ is the probing signal vector for each PA activation, and the corresponding received signal is given by
\begin{equation}
    \label{probingreceivesignal}
    y^{\text{prob}}_{k}[t]= \delta_{l,k}\mathbf{h}^H_k[t] \mathbf{W}^{\text{prob}}\sqrt{\alpha_{k}}\mathbf{s}^{\prime}_{k}[t] + z_k, \forall t \in \mathcal{T}.
\end{equation}

\subsubsection{Communication Model}
In the downlink MIMO communications, since all waveguides are driven by a common baseband signal, this signal is then converted by the RF chain and fed to the waveguides for radiation. Transmit beamforming is applied in the baseband to achieve spatial multiplexing. Let $s_k$ denote the transmit signal emitted by the PASS system toward all users, which can be expressed as:
\begin{equation}
    \label{transmitsignal}
    s=\sqrt{\alpha_k}{s_k},
\end{equation}
where transmit signal $s_k \in \mathbb{C}$ for user $k$ is the zero-mean unit-variance stationary process satisfying $\mathbb{E}\left[s_k^H{s_k}\right]=1$. Let the received signal at UE $k$ through channels manipulated by PASS, expressed as
\begin{equation}
    \label{receivedsignal}
    \begin{aligned}
        y_{k} = \underbrace{\delta_{l,k}{\mathbf{h}_{k}^H\mathbf{G}}\mathbf{w}_k\sqrt{\alpha_{k}}s_k}_{\text{desired signal}}+\underbrace{\delta_{l,k}\mathbf{h}_{k}^H\mathbf{G}\mathbf{w}_k\!\sum_{i=1,i\neq k}^{K}\!\sqrt{\alpha_i}s_i}_{\text{multi-user interference}}+z_k,
    \end{aligned}
\end{equation}
where $z_k \in \mathcal{CN}(0,\sigma^2)$ is the additive white Gaussian noise at user $k$, $\sigma^2$ is the noise power. Moreover, $\mathbf{w}_k \in \mathbb{C}^{N \times 1}$ denotes the transmit beamforming vector for user equipments (UEs), and pinching beamforming matrix denotes $W \in \mathbb{C}^{N \times K}$. The effective pinching beamforming can be formulated as $\mathbf{h}^H_{k}\mathbf{G}$. Hence, the signal-to-interference-plus-noise ratio (SINR) for user $k$ to decode its own signal can be expressed as
\begin{equation}
    \label{SINR}
        \mathrm{SINR}_k(\mathbf{W},\mathbf{X}) = \frac{\delta_{l,k}\left|\mathbf{h}_k^{H}\mathbf{G}{\mathbf{w}_k}\right|^{2}\alpha_k}{\delta_{l,k}\left|\mathbf{h}_k^{H}\mathbf{G}{\mathbf{w}_i}\right|^{2}\!\sum\limits_{i=1,i\neq k}^{K}\!\alpha_i+\sigma^{2}}, \forall k \in \mathcal{K},
\end{equation}
then, the corresponding achievable rate is
\begin{equation}
    \label{rate_k}
    R_k=\log_2\left(1+\mathrm{SINR}_k(\mathbf{W},\mathbf{X})\right),
\end{equation}
and the system sum rate can be expressed as $\sum_{k \in \mathcal{K}}\log_2 \left(1+\mathrm{SINR}_k(\mathbf{W},\mathbf{X})\right)$.

During the beam training, we train the LLM to get the selected pinching beamforming codeword from the generation codebook. Then, we utilize MRT technology to derive the optimal transmit beamforming in single-user case, which can be given by
\begin{equation}
    \label{MRT_w}
    \mathbf{w}^{\text{MRT}}_k(\mathbf{x}) = \sqrt{P_{\text{max}}}\frac{\mathbf{h}_k^{H}(\mathbf{x})}{\|\mathbf{h}_k^{H}(\mathbf{x})\|}.
\end{equation}

For multi-user case, to evaluate the accuracy of beam prediction and selection, we introduce the Top-$S$ function \cite{zheng2025beamllm, alrabeiah2020deep} to calculate the beam selection accuracy of training labels in candidate pinching beamforming sets within the top $S=\{1,3\}$ predicted beams ranked by model confidence. Formally, it is computed as
\begin{equation}
    \mathcal{TOP}_S = \frac{1}{N_S} \sum_{j=1}^{N_s} {1}_{\{m_j \in \mathcal{Q}_S\}},
    \label{topA}
\end{equation}
where $N_s$ denotes the total number of test samples, $m_j$ is the index of the ground truth optimal beam for $j$-th sample, and $\mathcal{Q}_S$ is index set for the Top-$S$ predicted beams. Then, we utilize the MMSE strategy to derive the optimized transmit beamforming with the selected candidate pinching beamforming sets from codebook generation, which can be expressed as
\begin{equation}
    \label{MMSE_w}
    \mathbf{W}^{\text{MMSE}} = \mathbf{H}^H(\mathbf{X})(\mathbf{H}(\mathbf{X})\mathbf{H}^H(\mathbf{X})+\sigma^2\mathbf{I}_k)^{-1}\sqrt{\mathbf{P}},
\end{equation}
where $\mathbf{P}$ represents the power allocation diagonal matrix and $\mathbf{I}_k$ is $k$-order identity matrix.

\section{Beam Training of LLM-Based PASS for Single-User Case}

In this section, we propose an LLM-based codebook generation and beam training mechanism based on LLM. We formulate the pinching beamforming codebook generation problem to maximize the beamforming gain. With the pinching beamforming generated by LLM-based codebook generation and labelled in the candidate pinching beamforming sets, the optimal transmit beamforming is derived by MRT technique. Moreover, we analyze the performance of LLM-based PASS.

\begin{figure}[t]
    \centering
    \includegraphics[width=3.6in]{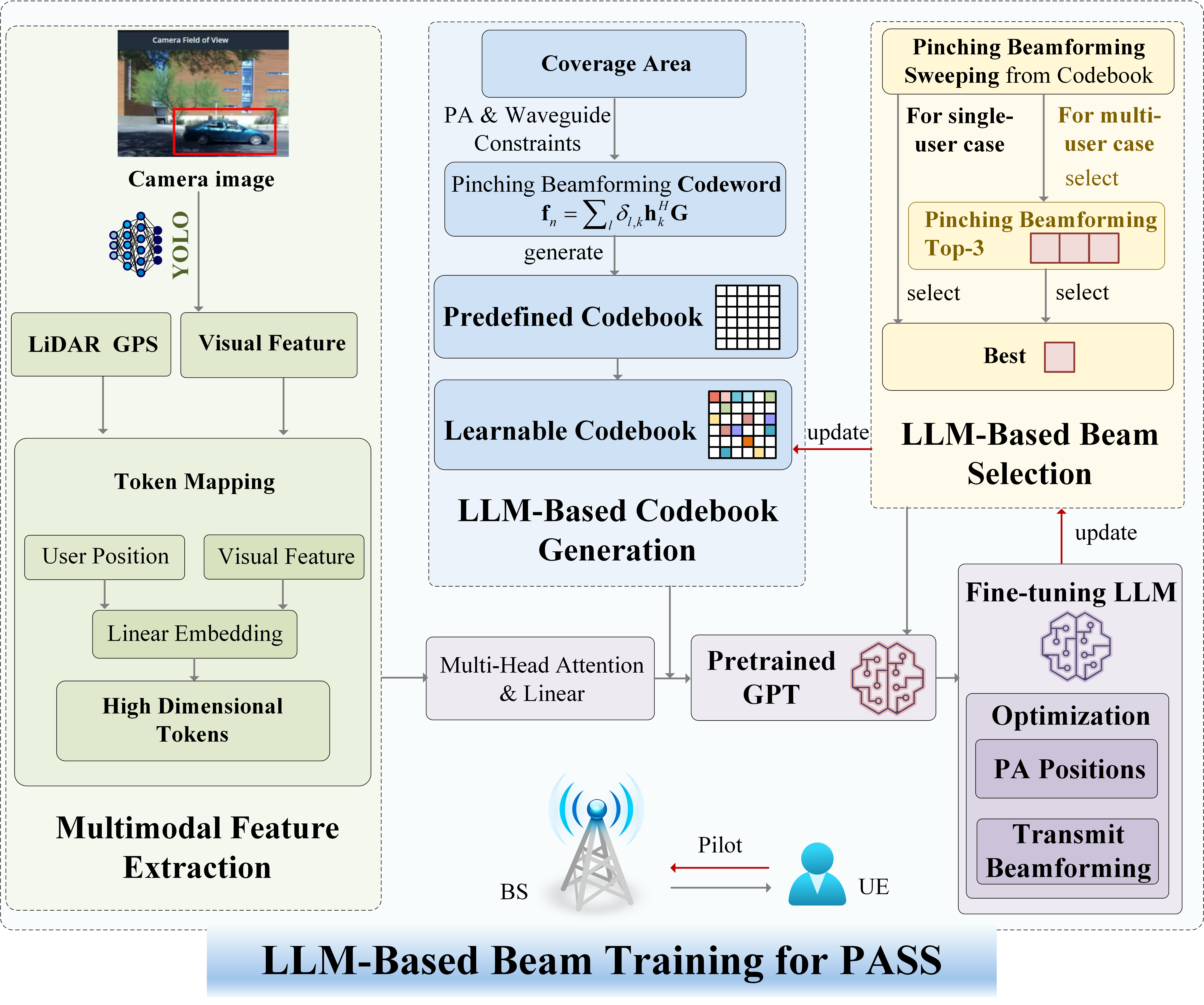}
    \caption{The proposed LLM-based Beam Training for PASS.}
    \label{LLM}
\end{figure}

As perfect channel state information (CSI) is unavailable, beam training is an effective approach to identify and predict the best pinching beamforming. However, for systems employing a narrow beam codebook, the corresponding training overhead can become substantial, and the probability of successfully selecting the optimal beam diminishes, particularly under low SNR conditions in single-user case. Given that beam selection at both the transmitter and receiver is highly dependent on the transceiver's surrounding environment, this work exploits multimodal sensing information at the BS to enhance beam selection trained by LLM and to develop a multimodal-assisted beam training framework. This beam training framework is shown as Fig. \ref{LLM}.

\subsection{Problem Formulation}

Single-user case is a special case in multi-user case, the received signal for the user can be simplified as
\begin{equation}
    \label{receivedsignal1}
    y_1 = \delta_{1,1}\mathbf{h}_1^{H}\mathbf{G}{\mathbf{w}_1}\sqrt{\alpha_1}{s_1},
\end{equation}
we focus on the single-user case where $L$ PAs are deployed along a waveguide. Suppose that the $L$ PAs are positioned at $\mathbf{x}_{1}=[x_{1,1},x_{1,2},...,x_{1,L}]^T \in \mathbb{R}^{L \times 1}$ along the waveguide $1$, and the single user is at location ${\bm{\psi}}^{\text{U}}_1=[x_1,y_1,0], k = K =1$, where $0 \le x_1 \le x^{\max}$ and $0 \le y_1 \le y_N$. The downlink channel vector is expressed as
\begin{equation}
    \label{channelvector1}
    h^H_{1,l,1} (x_{1,l}) = \frac{\eta e^{-i{\kappa}\left|{\bm{\psi}}_1^{\text{U}}-{\bm{\psi}}_{1,l}^{\text{PA}}\right|}}{\left|{\bm{\psi}}_1^{\text{U}}-{\bm{\psi}}_{1,l}^{\text{PA}}\right|}, 
\end{equation}
where $\kappa=\frac{2\pi}{\lambda}$ is the wave-domain number. Hence, the channel vector of the PAs along wavegudie $1$ is denoted by $\mathbf{h}^H_{1,1} (\mathbf{x}_1) \in \mathbb{C}^{1 \times L}$.

For single-user case, the system sum rate is reduced to the form ignoring multi-user interference. The signal-to-noise ratio (SNR) for the only one user to decode its signal is given by
\begin{equation}
    \label{SNR1}
    \mathrm{SNR}_1 = \frac{\delta_{l,1}\left|\mathbf{h}_1^{H}\mathbf{G}{\mathbf{w}_1}\right|^{2}\alpha_1}{\sigma^{2}},\text{ } k=K=1,
\end{equation}
and the data rate of user is as follows,
\begin{equation}
    \label{datarate1}
    R_1 = \log_2\left(1+\mathrm{SNR}_1\right).
\end{equation}

To enhance the accuracy of beam selection and prediction, we employ the beamforming matching strength $\|\mathbf{f}_n(\mathbf{x}_n){\mathbf{w}_k}\|^{2}$ that evaluates the predicted beam alignment with the direction of channel, referred to as beamforming gain. Therefore, maximizing the system sum rate can be recast as maximizing the beamforming gain under the MRT transmit beamforming strategy (\ref{MRT_w}). We formulate the beamforming gain maximization problem by searching for the optimal pinching beamforming $\mathbf{x}_n^{*}$,
\begin{subequations}
    \label{max_gain0}
    \begin{align}
            \mathbf{P}_{0.0}: \quad & \arg \max_{\mathcal{F}(\cdot),\mathcal{D}_{\theta}(\cdot)}\|\mathbf{f}_n(\mathbf{x}_n){\mathbf{w}_k}\|^{2},
            \label{max_beamforminggain}\\
            \mathrm{s.t.~} \quad & \hat{\mathbf{x}}_n = \mathcal{F}(y_{1:T},\mathbf{F}_{1:T},\widehat{B}), \forall l\in\mathcal{L}, k=K=1, n\!=\!N\!=\!1,
            \label{PAposition_select0}\\
            & \mathbf{F}_t = \mathcal{D}_{\theta}(\{\widehat{B}[t]\}^{t}_1), \forall t \in \mathcal{T}=\{1,\dots,T\},
            \label{codebookgeneration_1}
    \end{align}
\end{subequations}
where $\hat{\mathbf{x}}_n$ represents the best positions selected from the training labels in the selected pinching beamforming set for the user, $y$ denotes the received signal during the channel probing $t \in \mathcal{T}=\{1,\dots,T\}$, $\widehat{B}$ is the high-dimensional tokens fed into the LLM, $\mathcal{F}$ denotes the specific mapping during the beam training based on LLM, predicting the optimal positions of PAs based on the signal, high-dimensional tokens and generated codebook, and $\mathcal{D}_{\theta}$ is the calculation process of the codebook generation based on the LLM with the learning parameter ${\theta}$. (\ref{PAposition_select0}) constraints the optimal positions of PAs are predicted and selected from the ideal pinching beamforming set for the user, and (\ref{codebookgeneration_1}) means the pinching beamforming codeword is processed by first LLM for codebook generation.
For simplicity, the problem (\ref{max_gain0}) is reduced to
\begin{subequations}
    \label{max_0}
    \begin{align}
            \mathbf{P}_{0.1}: \quad & \min_{\mathcal{F}_{\theta}(\cdot),\mathcal{D}_{\theta}(\cdot)}\|\mathbf{x}^*_n-\hat{\mathbf{x}}_n\|^{2},
            \label{min_distance}\\
            \mathrm{s.t.~} \quad & \text{(\ref{PAposition_select0}), (\ref{codebookgeneration_1})}, \forall l\in\mathcal{L}, k=K=1, n\!=\!N\!=\!1.
    \end{align}
\end{subequations}

\subsection{Algorithm Design}

\subsubsection{Multimodal Feature Extraction}
The multimodal feature extraction is the first module of LLM-enabled PASS, which is responsible for extracting the dynamic multimodal information of the target user from the cameras deployed at the BS and converting it into a time-sequential structured representation for subsequent LLM processing. The target detector, e.g., YOLO series \cite{sohan2024YOLOv8}, outputs the user's bounding box of the image in each time slot, which is converted into a fixed-dimension token sequence through normalization, patch partitioning, and linear embedding. Specifically, we extract multimodal features from camera images, LiDAR, and GPS to construct the high-dimensional tokens for codebook generation and beam selection that leverages the strong generalization and inference abilities of pretrained LLMs.

To reduce the overhead and latency of conventional beam training, we incorporate real-time multimodal feedback from cameras deployed at the BS. The goal of this module is to extract the user's spatio-temporal movement from raw video frames and transform it into a structured token sequence that is compatible with language models. Specifically, a multimodal detector, such as YOLO, is used to extract the bounding box of the target user in each time slot. Let the bounding box at the $t$-th time slot be denoted as
\begin{equation}
    \label{boundingbox}
    \mathbf{b}_{k,t} = [x_{k,t} \text{ }  y_{k,t} \text{ } LT^{\text{wid}}_{k,t} \text{ } LT^{\text{hei}}_{k,t}]^T \in \mathbb{R}^4,
\end{equation}
where $x_{k,t}=x_k$ and $y_{k,t}=y_k$ denote the center coordinates of the detected bounding box, $LT^{\text{wid}}_{k,t}$ and $LT^{\text{hei}}_{k,t}$ are the width and height of the detected box. Let the bounding box series of user $k$ be represented by $\mathbf{B}_k \in \mathbb{R}^{4\times T}$ over $T$ time slots, where each row corresponds to one of the four bounding box attributes, which is represented as
\begin{equation}
    \label{boundingboxseries}
    \mathbf{B}_k = [\mathbf{b}_{k,1},\mathbf{b}_{k,2},\dots,\mathbf{b}_{k,T}] \in \mathbb{R}^{4\times T}.
\end{equation}

For single-user case, the user position and multimodal features are extracted via the multimodal detector, i.e., YOLO, yielding bounding boxes $\mathbf{b}_{1,t}$ to generate the bounding box series of this user $\mathbf{B}_1=[\mathbf{b}_{1,1},\dots,\mathbf{b}_{1,t},\dots,\mathbf{b}_{1,T}] \in \mathbb{R}^{4\times T}$. After token mapping, the combined information is represented as high-dimensional tokens $\widehat{\mathbf{B}}_P^{(i)}, \forall i=\{1,2,3,4\}$.
To bridge the modality gap between visual sequences and the LLM input format, we design a token mapping part that transforms raw spatio-temporal bounding box sequences into high-dimensional tokens. These tokens are aligned in structure and semantics with the textual input space, allowing seamless integration with the attention-based reasoning pipeline of LLM.

To eliminate scale shifts and normalize across different scenes, we apply reversible instance normalization (RevIN) across each dimension,
\begin{equation}
    \label{boundingbox_norm}
    \widetilde{\mathbf{B}}^{(i)}_k = \frac{\mathbf{B}^{(i)}_k-\mu^{(i)}_B}{\sigma^{(i)}_B}, \text{ } \forall i=\{1,2,3,4\},
\end{equation}
where $\mu^{(i)}_B$ is the mean of $\mathbf{B}_k$, $\sigma^{(i)}_B$ is the standard deviation of $\mathbf{B}_k$. RevIN further stores $(\mu^{(i)}_B, \sigma^{(i)}_B)$ for subsequent inverse transformation if needed during inference. 

Since LLMs cannot directly process the multimodal features and natural language simultaneously. Subsequently, the bounding box sequences need to be segmented into patches and converted into high-dimensional tokens using learnable linear projections, aligning multimodal information semantically with textual-like tokens for effective inference by the LLM. The series $\widetilde{\mathbf{B}}^{(i)}_k$ are segmented into overlapping or non-overlapping patches of length $L^{\text{patch}}$, using a temporal sliding window of step size $L^{\text{full}}$. The total number of patches for each channel is given by
\begin{equation}
    \label{num_patch}
    N_P=\lfloor\frac{T-L^{\text{patch}}}{L^{\text{full}}}\rfloor + 1.
\end{equation}

We denote the patch matrix from $i$-th attribute as $\widetilde{\mathbf{B}}^{(i)}_P \in \mathbb{R}^{N_P\times L^{\text{patch}}}$. To convert the raw patch data into a high-dimensional embedding suitable for LLM-based reasoning, each patch is passed through a learnable linear projection to obtain high-dimensional tokens of patch in $s^{\text{patch}}$-th dimension space, which can be expressed as
\begin{equation}
    \label{learnable_projection}
    {\widehat{\mathbf{B}}}^{(i)}_P = \widetilde{\mathbf{B}}^{(i)}_P \cdot \mathbf{EM}^{(i)} + \mathbf{m}^{(i)}, \widehat{\mathbf{B}}^{(i)}_P \in \mathbb{R}^{P\times s^{\text{patch}}},
\end{equation}
where $\mathbf{EM}^{(i)}\in \mathbb{R}^{L^{\text{patch}}\times s^{\text{patch}}}$ and $\mathbf{m}^{(i)} \in \mathbb{R}^{s^{\text{patch}}}$ are learnable parameters.

\subsubsection{LLM-Based Codebook Generation}
In conventional massive MIMO systems, the discrete fourier transform (DFT)-based codebooks have been widely adopted due to their mathematical simplicity and uniform spatial coverage. These codebooks are constructed solely based on the antenna array geometry, assuming ideal linear structures and uniform phase progression across elements. While such predefined beamforming vectors provide tractable solutions for static, symmetrical deployments, they exhibit fundamental limitations in flexible and adaptable PASS.

We propose an LLM-based codebook generation mechanism that leverages powerful multimodal alignment capabilities to dynamically synthesize pinching beamforming codewords tailored to the dynamic communication scenarios. Considering the coupled positions of PAs and CSI, we design an adaptive codebook generation and beam selection mechanism, which first declares the initial codebook consists of all pinching beamforming, and then, updates the initial one by the optimized PA positions and latest user locations to generate a learnable codebook. Specifically, the predefined codebook $\mathbf{F}$ is obtained by the pinching beamforming according to (\ref{pinchingbeamforming}). 
Based on (\ref{pinchingbeamforming}), codeword in the $t$-th time slot can be formulated as
\begin{equation}
    \label{codeword_tau}
    \mathbf{f}_n[\mathbf{X},t] = \sum_l\delta_{l,k}[t](\mathbf{h}^H_k[\mathbf{X},t]\mathbf{G}[\mathbf{X},t]), \forall t \in \mathcal{T}=\{1,2,\dots,T\}.
\end{equation}

We can further derive $\mathbf{F}[t]=\{\mathbf{f}_n[t]\}^{N}_{n=1}$ from $\mathbf{f}_n \gets \mathcal{D}_{\theta}(\widehat{B}^{(i)}_P)$. Through the codebook generation based on LLM, we can obtain the learnable codebook $\widetilde{\mathbf{F}}$, consists of iterative updated pinching beamforming codewords, given by $\widetilde{\mathbf{F}}=\left\{\mathbf{f}_1[T],\mathbf{f}_2[T],\dots,\mathbf{f}_N[T]\right\}$. And the optimal codeword $\mathbf{f}^*_n$ and optimal pinching beamforming can be obtained after convergence. With the optimal pinching beamforming, the optimal transmit beamforming can be derived with the MRT technique according to (\ref{MRT_w}), which is given by
\begin{equation}
    \label{optimal_tranmitw1}
    \mathbf{w}^{\text{MRT}}_1 = \sqrt{P_{\text{max}}}\frac{\mathbf{h}_1^{H}(\mathbf{x}_1^*)}{\|\mathbf{h}_1^{H}(\mathbf{x}_1^*)\|}.
\end{equation}

\subsubsection{Performance Analysis}
For single-user case, the reliability of the proposed LLM-enabled PASS is evaluated. 
\begin{proposition}
    For single-user case, the outage probability is achievable at a high SNR by LLM-enabled PASS, which is smaller than that of the conventional massive MIMO system.
\end{proposition}

\begin{IEEEproof}
    With a given maximum achievable rate $R_1^{\max}$ of user $1$, define corresponding SNR threshold $\epsilon_1 = 2^{R_1^{\max}} - 1$. The outage probability is 
    \begin{equation}
        \label{outage_1}
        E^{\text{outage}} = \mathbb{E}\left(R_1 < R_1^{\max}\right) = \mathbb{E}\left(\mathrm{SNR}_1 < \epsilon_1\right),
    \end{equation}
    further formulated as
    \begin{equation}
        \label{outage_11}
        E^{\text{outage}}  = \mathbb{E}(\delta_{1,1}=0) + \mathbb{E}\left(\left| \mathbf{h}_1^{H} \mathbf{G} \mathbf{w}_1 \right|^2 < \epsilon_1 \sigma^2,~\delta_{1,1}=1\right).
    \end{equation}

    Assume that the LoS probability decays exponentially with the distance between the user and the PA, i.e.,
    \begin{equation}
        \label{delta11_outage}
    \mathbb{E}(\delta_{1,1}=1) = e^{\left(-\varphi_1 \min_{l}\|\bm{\psi}_1^{\text{U}} - \bm{\psi}_{1,l}^{\text{PA}}\|\right)},
    \end{equation}
    where $\varphi_1$ is the blockage density parameter for a single user. Define $d^{\min}(y_1) = \min_{l} \| [x_1, y_1, 0] - [x_{1,l}, 0, 0] \|$.
    For a user uniformly distributed along $y_1 \in \left[0 , y^{\max}\right]$, the average outage probability is
    \begin{equation}
        \label{outage_111}
        \begin{aligned}
            E^{\text{outage}} = &~\frac{1}{y^{\max}} \int_{0}^{y^{\max}} \left[ 1 - e^{-\varphi_1 d^{\min}(y_1)} \right] dy_1 \\
            &+ \frac{1}{y^{\max}} \int_{S} e^{-\varphi_1 d^{\min}(y_1)} dy_1,
            \end{aligned}
    \end{equation}
    where the integration set $S = \left\{ y_1 : d^{\min}(y_1) \geq \tau_1 \right\}$, with $\tau_1 = \sqrt{\frac{\eta_1 P^{\max}}{\epsilon_1 \sigma^2}}$. When $P^{\max}$ is sufficiently large, $\tau_1 \to 0$, so the second term of (\ref{outage_111}) vanishes, and the outage probability simplifies to
    \begin{equation}
        \label{surrogate_outage111}
        E^{\text{outage}} \approx 1 - \frac{1}{y^{\max}} \int_{0}^{y^{\max}} e^{-\varphi_1 d^{\min}(y_1)} dy_1.
    \end{equation}

    As for conventional massive MIMO, a fixed antenna at position $\psi_{A}$,
    \begin{equation}
        \label{MIMO_outage1}
        E^{\text{outage}}_{\text{conv}} \approx 1 - \frac{1}{y^{\max}} \int_{0}^{y^{\max}} e^{-\varphi_1 \| [x_1, y_1, 0] - \psi_C \|} dy_1.
    \end{equation}
    
    Since $d^{\min}(y_1) \leq \| [x_1, y_1, 0] - \psi_C \|$ always holds with PA movement, it follows that
    \begin{equation}
        \label{compare_outage}
        E^{\text{outage}} < E^{\text{outage}}_{\text{conv}},
    \end{equation}
    demonstrates that LLM-enabled PASS has the advantages that support flexible positions of PAs and dynamic movement of users simultaneously, leading to strictly lower outage probability compared to conventional massive MIMO.

    The above proof provides theoretical evidence for the reliability gain of the proposed LLM-enabled PASS. Thus, the proposed LLM-enabled PASS outperforms conventional massive MIMO in terms of reliability.
\end{IEEEproof}

\begin{lemma}
    For the single-user case, under the blockage model of the single-waveguide PASS configuration, assuming the user is uniformly distributed along $y_1 \in [0, y^{\max}]$ and the PA can be optimally placed to minimize the distance to the user, the outage probability at high SNR is given by (\ref{surrogate_outage111}).
\end{lemma}
    
\begin{IEEEproof}
    The outage probability is defined as (\ref{outage_1}) and decomposed as (\ref{outage_11}).
    The probability of LoS, i.e., $\delta_{1,1}=1$, is modeled by exponential decay with the closest distance between PA and user as (\ref{delta11_outage}).
    For a user at $[x_1, y_1, 0]$, the minimum distance is $d^{\min}(y_1)$ due to optimal PA positions.
    At high SNR ($E^{\max} \to \infty$), the SNR threshold $\tau_1 = \sqrt{ \frac{ \eta_1 E^{\max} }{ \epsilon_1 \sigma^2 } } \to \infty$, making it almost always possible to meet the SNR constraint whenever $\delta_{1,1}=1$. Thus, the second term in (\ref{outage_11}) vanishes and only the blockage event remains:
    \begin{equation}
        E^{\text{outage}} \approx \mathbb{E}_{y_1}\left[ 1 - e^{-\varphi_1 d^{\min}(y_1)} \right].
    \end{equation}
    
    Taking the expectation over the uniform distribution of $y_1$ yields (\ref{highSNR_outage}), 
    \begin{figure*}[t]
        \begin{equation}
            \label{highSNR_outage}
            \begin{aligned}
                E^{\text{outage}} & = 1 - \frac{1}{y^{\max}} \int_{0}^{y^{\max}} e^{-\varphi_1 d^{\min}(y_1)} dy_1 + \frac{1}{y^{\max}} \int_{0}^{\tau_2} e^{-\varphi_1 d^{\min}(y_1)} dy_1 + \frac{1}{y^{\max}} \int_{\tau_3}^{y^{\max}} e^{-\varphi_1 d^{\min}(y_1)} dy_1 \\
                & \approx 1 - \frac{1}{y^{\max}} \int_{0}^{y^{\max}} e^{-\varphi_1 d^{\min}(y_1)} dy_1,
            \end{aligned}
        \end{equation}
    \end{figure*}
    where $\tau_2 = \max\{0,\sqrt{{\tau_1}^2-{d^{\min}}^2}\} $ and $\tau_3 = \min\{\sqrt{{\tau_1}^2-{d^{\min}}^2},y^{\max}\}$. As $\text{SNR}_1$ is high, $\tau_2 \to 0$ and $\tau_3 \to y^{\max}$, the above equation is dominated by the second term, the third and fourth terms close to 0. Therefore, PASS can achieve the outage probability at the high SNR.
\end{IEEEproof}

Compared to the multi-user case, the single-user case substantially reduces computational complexity, as there is no need to consider multi-user interference. The entire training process can produce optimal locations of PAs and optimal pinching beamforming with a single LLM inference, supporting real-time beam training in practical deployments. The computational complexity of the proposed LLM-based beam training is $\mathcal{O}(TN_{\text{batch}}\left[N_P(s^{\text{patch}})^2 + L\right])$.

\section{Beam Training of LLM-Based PASS for Multi-User Case}

In this section, we propose a multimodal-assisted beam training framework for PASS based on LLM with supervised end-to-end training. First, we design a multimodal feature extraction module to retrieve the information of the camera image, LiDAR, and GPS. Then, we generate an LLM-based codebook to select the best pinching beamforming for users and save it as the candidate pinching beamforming sets, which has the same process as the mentioned in {\bf{Section III}}. We formulate the LLM-based pinching beamforming codebook generation and beam selection problem under system sum rate maximization problem to jointly optimize pinching beamforming and transmit beamforming. To construct the training samples for beam training, the above problem is decomposed into two subproblems, i.e., Top-$S$ selection problem and rate maximization problem. With the best selection from candidate sets, we further utilize the MMSE transmit beamforming strategy to explore the best transmit beamforming for users. For multi-user case, the proposed LLM-enabled beam training for PASS is also shown as Fig. \ref{LLM}.

\subsection{Problem Formulation}

To enhance probing performance and accurately capture channel characteristics under multi-user interference, digital beamforming is incorporated during the channel probing stage, enabling more effective signal separation and improved evaluation of candidate pinching beamforming configurations. For multi-user case, since PAs are flexible along waveguides, the locations of PAs could significantly affect achievable data rates of users in the MIMO communication scenario. With the given ideal pinching beamforming codebook $\mathbf{F}^*$, the LLM-based pinching beamforming codebook generation and beam selection problem under the system sum rate can be formulated as
    \begin{subequations}
        \label{max_sumrate1}
        \begin{align}
                \mathbf{P}_{1.0}: \quad & \max_{\mathcal{F}_{\theta}(\cdot),\mathcal{D}_{\theta}(\cdot)}\sum_{k \in \mathcal{K}}\log_2 \left(1+\mathrm{SINR}_k(\mathbf{W},\mathbf{X})\right),
                \label{maxsumratek}\\
                \mathrm{s.t.~} \quad & R_{k} \ge R_{k}^{\mathrm{min}},k\in\mathcal{K},
                \label{rateconstraints}\\
                & 0<\alpha_{k}<1,\sum_{k=1}^{K}\alpha_{k}=1,k\in\mathcal{K}, 
                \label{powerconstraints1}\\
                & \sum_{k \in \mathcal{K}}\|\mathbf{w}_k\|^2_2 \le P_{\max},
                \label{powerconstraints2}\\
                & x_{n,l}-x_{n,l-1} \ge \Delta, l\in\mathcal{L}, n\in\mathcal{N},
                \label{PA_minx}\\
                & x_{n,l}\in[0,x_{n,L}], l\in\mathcal{L}, n\in\mathcal{N},
                \label{PA_x}\\
                & \hat{\mathbf{X}} = \mathcal{F}_{\theta}(y_{1:T},\mathbf{F}_{1:T},\widehat{B}^{(i)}_P,\mathbf{W}_{1:T}),
                \label{PAposition_select1}\\
                & \{\mathbf{F}_t,\mathbf{W}_t\} = \mathcal{D}_{\theta}(y_{1:T},\mathbf{F}_{1:T},\widehat{B}^{(i)}_P), \forall t \in \mathcal{T},
                \label{codebookgeneration_k}
        \end{align}
    \end{subequations}
    where constraint (\ref{rateconstraints}) guarantees the minimum data rate of each user, constraint (\ref{powerconstraints1}) ensures the power allocation coefficient constraint and variation range for all users, constraint (\ref{powerconstraints2}) denotes the maximum transmit power of BS $P_{\max}$, constraint (\ref{PA_minx}) denotes the minimum antenna space $\Delta$ to avoid mutual coupling, constraint (\ref{PA_x}) limits the location of each PA to the maximum length of corresponding waveguide, constraint (\ref{PAposition_select1}) limits the best positions of PAs predicted and selected from the ideal pinching beamforming sets for users by LLM-based beam selection $\mathcal{F}_{\theta}$, and constraint (\ref{codebookgeneration_k}) means the pinching beamforming and transmit beamforming is processed by first LLM generation LLM.
    Different from the $\mathcal{F}$ mapping of beam selection in {\bf{Section III}}, $\mathcal{F}_{\theta}$ is the parametric mapping with learning parameter $\theta$, because information of multiple users needs to be considered to mitigate inner interference among users exits in multi-user case.
    To further increase the sum rate, we aim to calculate the best transmit beamforming under MMSE strategy based on the best explored pinching beamforming $\hat{\mathbf{X}}$. 

\subsection{Algorithm Design}

\subsubsection{LLM Structure}

    The process begins with extracting and fusing multimodal features from camera images, LiDAR, and GPS data into high-dimensional tokens, which comprehensively characterize the surrounding environment.
    Leveraging these tokens, a supervised learning mechanism based on the first LLM is employed to generate the pinching beamforming codebook, which defines a sequence of candidate PA positions for effective spatial probing.
    The LLM-based beam selection based on the second LLM predicts the optimal pinching beam by mapping the structured tokens to beam indices, enabling robust adaptation to locations of users and channel state. Finally, given the predicted pinching beam, the MMSE transmit beamforming strategy is applied to determine the best transmit beamforming, thereby further enhancing beam training for users.

\subsubsection{LLM-Based Beam Selection}    
The LLM-based beam selection module forms the core of the LLM-enabled PASS, bridging high-level token inference with low-level physical reconfiguration. This module is designed to select the best predicted beam configurations by leveraging spatial-contextual information of tokens extracted by the pretrained LLM from camera images, historical locations of PAs, and users.
To predict the best beam, the last module of LLM-enabled PASS is LLM-based beam selection. The core idea behind this module is to couple high-level token inference with low-level physical layout reconfiguration of PASS. Specifically, a pretrained LLM learns the extracted spatial and scene-aware priors that enable the prediction of the selected beam based on user distribution, image inputs, and PA configurations. Meanwhile, the PA position $\bm{\psi}^{\text{PA}}_{n,l}(\mathbf{X})$ is treated as a tunable physical variable that can be optimized to improve beam alignment and SINR. Different from traditional methods that rely on direct channel measurements or exhaustive beam training, we propose an LLM approach to predict the best beam directly using the pretrained LLM to tackle the non-convex $\mathbf{P}_{1.0}$, significantly reducing the overhead. From $\mathbf{P}_{1.0}$, the best codeword $\mathbf{f}^*_n$ in the selected candidate pinching beamforming sets are necessary to maximize the beamforming gain, which can be denoted by
\begin{equation}
    \label{gain}
    \mathbf{F}_{\mathrm{sel}} = \arg \max \|\mathbf{f}^*_n(t){\mathbf{w}_k}\|^{2},
\end{equation}
where $\mathbf{w}_k$ is transmit beamforming for user $k$, dynamically adapted by instantaneous CSI.
To address the above challenge, we construct the training samples by decomposing $\mathbf{P}_{1.0}$ in to two subsections. Since obtaining the optimal solution is intractable
    \begin{equation}
        \label{optimal_xw}
        \{\mathbf{X}^*,\mathbf{W}^*\} = \arg \max_{\mathcal{F}_{\theta}(\cdot),\mathcal{D}_{\theta}(\cdot)}\sum_{k \in \mathcal{K}}\log_2 \left(1+\mathrm{SINR}_k(\mathbf{W},\mathbf{X})\right),
    \end{equation}
    we reformulate the problem as a supervised learning problem, allowing the model to learn effective beamforming strategies from labelled data. The supervised learning training problem can be decomposed into two subproblems, i.e., Top-$S$ selection problem and rate maximization problem.
    
    The Top-$S$ selection problem aims to find the candidate pinching beamforming sets, which can be formulated as $\{\mathbf{F}^{\text{c}},\mathbf{W}^{\text{c}}\} = \arg \mathcal{TOP}_S(\|\mathbf{f}_n(\mathbf{x}_n){\mathbf{w}_k}\|^2)$. Then, with the explored $\{\mathbf{F}^{\text{c}},\mathbf{X}^{\text{c}}\}$ candidate pinching beamforming sets, we formulate the rate maximization problem, which can be reduced to
    \begin{subequations}
        \label{max_1}
        \begin{align}
                \mathbf{P}_{1.1}: \quad & \min_{\mathcal{F}_{\theta}(\cdot)}\|\mathbf{X}^*-\hat{\mathbf{X}}\|^{2},
                \label{min_distancek}\\
                \mathrm{s.t.~} \quad & \text{(\ref{rateconstraints})-(\ref{codebookgeneration_k})}.
        \end{align}
    \end{subequations}

We perform an end-to-end supervised training strategy for beam training process. First, we train the adapter modules of pretrained LLM while freezing original LLM parameters, enabling initial alignment between the multimodal-channel features and the high-dimensional mapping space for LLM.
Next, we fine-tune the LLM using the MoE-LoRA, dynamically selecting subsets of parameters to efficiently capture tokens across multi-user communication scenarios. This two-step training significantly enhances beam selection capability and reduces the computational complexity.
High-dimensional tokens ${\widehat{\mathbf{B}}}^{(i)}_P$ and updated codeword $\mathbf{f}_n$ codebook $\widetilde{\mathbf{F}}$ are configured as structured tokens to reflect both spatial and semantic multimodal cues by multi-head attention and linear layers, which are the inputs to the pretrained LLM, denoted by
\begin{equation}
    \label{input_preLLM}
    \mathcal{X} = \text{Softmax}\left[\text{Adapter}(\left\{\{{\widehat{\mathbf{B}}}^{(i)}_P\},\mathbf{F}_{1:T}\right\})\right],
\end{equation}
where $\text{Softmax}[\cdot]$ function maintains robust of the structured tokens, $\text{Adapter}(\cdot)$ is the combined function of multi-head and linear layers. The outputs of pretrained LLM are obtained by $\mathbf{O}(\mathcal{X})$ with the pretrained weight parameter $a_0$. 

Further, the pretrained LLM is fine-tuned using a mixture of experts with low-rank adaptation (MoE-LoRA) to capture the relationships within these inputs effectively, which indicates the specialized alignment layers designed to match the semantic output of LLM to the beam selection task. The standard LoRA layers aim to get low-rank matrices, e.g., positions of PAs and transmit beamforming for users, which is tailored for joint pinching beamforming and transmit beamforming optimization. Specifically, we define the output probability distribution of the beam via LoRA layers as
\begin{equation}
    \label{LoRA}
    \mathbb{P}(n,l,k|\mathbf{O}(\mathcal{X});\Theta) = \text{LoRA}(\text{Transformers}(\mathbf{O}(\mathcal{X});\Theta)),
\end{equation}
where $\text{Transformers}(\cdot)$ function represents the inputs fine-tuned by the transformer layers, $\text{LoRA}(\cdot)$ function is the simplified-form function of beam selection via LoRA layers, $\Theta=\{\theta_0,\theta_1,\theta_2,\dots,\theta_k,\dots,\theta_K\}$ represents the parameter set of the fine-tuning LLM. To further predict optimal beam, the MoE layers are utilized to learn beam training and selection parameters. A gating network is then employed to select and combine different experts for beam training and selection, facilitating an aggregation mechanism for experts. The optimal predicted beam distribution probability can be written as (\ref{optimalbeamprediction}),
\begin{figure*}[t]
    \begin{equation}
        \label{optimalbeamprediction}
        \mathbb{Z}(n,l,k|\mathbf{O}(\mathcal{X});\Theta) = a_0 \mathbb{P}(n,l,k|\mathbf{O}(\mathcal{X});\Theta) + \frac{\text{rank}(\mathbf{X})}{\eta_{\text{MoE}}}\sum_{k \in \mathcal{K}}a_k{\text{MoE}(\mathbf{h}^H_k\mathbf{G}\mathbf{w}_k)}\mathbb{P}(n,l,k|\mathbf{O}(\mathcal{X});\Theta),
    \end{equation}
\end{figure*}
where $\text{MoE}(\cdot)$ function is the simplified-form distribution probability function of via MoE layers, $\text{rank}(\mathbf{X})$ is the rank of $\mathbf{X}$, $\eta_{\text{MoE}}$ is hyperparameter of rank of $\mathbf{X}$. 

For the rate maximization problem, the cross-entropy loss function is employed, which ensures the trade-off between training and computational complexity, and enhances the beam selection capability for PASS. This cross-entropy loss function can simultaneously optimize beam selection accuracy and PA position refinement during the LLM-based beam training for PASS, which is defined as follows.
\begin{equation}
    \label{loss_crossentropy}
\text{LOSS}_{\text{LLM}} = \sum_{k \in \mathcal{K}} \theta_k \text{loss}_{n}(\mathbb{Z}),
\end{equation}
where $\text{loss}_{k}$ denotes the specific loss function for user $k$, linearly combined with corresponding weight $\theta_k$. To balance the training across beam selection and selection effectively, we use the dynamic weight average algorithm to dynamically adjust every weight at every training epoch. With the best pinching beamforming, we use the MMSE to obtain the best transmit beamforming based on (\ref{MMSE_w}), given by
\begin{equation}
    \label{best_w}
    \hat{\mathbf{W}}(\hat{\mathbf{X}}) = \mathbf{H}^H(\hat{\mathbf{X}})(\mathbf{H}(\hat{\mathbf{X}})\mathbf{H}^H(\hat{\mathbf{X}})+\sigma^2\mathbf{I}_k)^{-1}\sqrt{\mathbf{P}}.
\end{equation}

The LLM-based optimal beam training algorithms for PASS are separated into two parts, i.e., inference and training, shown as Algorithm \ref{alg:llm_pass1} and Algorithm \ref{alg:llm_pass2}.

\begin{algorithm}[htbp]
    \caption{LLM-Based Beam Training Algorithm for PASS (Inference)}
    \label{alg:llm_pass1}
    \begin{algorithmic}[1]
    \REQUIRE Initial codebook $\mathbf{F}$, initial PA positions $\mathbf{X}^{(0)}$, multimodal embedding sequence $\{\widetilde{\mathbf{B}}^{(i)}_P\}$
    \ENSURE The best beam selection $\mathbf{F}_{\mathrm{sel}}$, the best PA positions $\hat{\mathbf{X}}$ and transmit beamforming $\hat{\mathbf{W}}$
        \STATE \textbf{Multimodal Feature Extraction:} 
        \STATE Extract multimodal features via token mapping: ${\widehat{\mathbf{B}}}^{(i)}_P = \widetilde{\mathbf{B}}^{(i)}_P \cdot \mathbf{EM}^{(i)} + \mathbf{m}^{(i)}, \forall i=\{1,2,3,4\}$
        \STATE \textbf{LLM-Based Codebook Generation:}
        \STATE Update codewords based on PA configurations obtain (\ref{codeword_tau})
        \STATE Transmit probing signal using the generated codebook and obtain (\Ref{probingreceivesignal})
        \STATE \textbf{LLM-Based Beam Selection:}
        \STATE Compute structured tokens for pretrained LLM: $\mathcal{X} = \text{Softmax}\left[\text{Adapter}(\left\{\{{\widehat{\mathbf{B}}}^{(i)}_P\},\mathbf{F}_{1:T}\right\})\right]$
        \STATE Obtain Top-$S$ candidate beams using (\ref{topA})
        \STATE Estimate $\hat{\mathbf{X}}$ from (\ref{PAposition_select1})
        \STATE For communication stage, obtain MMSE transmit beamforming $\hat{\mathbf{W}}$ by (\ref{best_w})
    \RETURN Best beam selection $\mathbf{F}_{\mathrm{sel}}$, PA positions $\hat{\mathbf{X}}$, and transmit beamforming $\hat{\mathbf{W}}$
    \end{algorithmic}
\end{algorithm}

\begin{algorithm}[htbp]
    \caption{LLM-based Beam Training Algorithm for PASS (Training)}
    \label{alg:llm_pass2}
    \begin{algorithmic}[1]
    \REQUIRE Same as Algorithm \ref{alg:llm_pass1}
    \ENSURE Same as Algorithm \ref{alg:llm_pass1}
        \FOR {each batch}
        \STATE Execute Algorithm \ref{alg:llm_pass1} for sampling
        \STATE \textbf{MoE-LoRA Fine-tuning:} Update layers using cross-entropy loss for beam selection, regression for PA location
        \STATE $\mathbb{P}(n,l,k|\mathbf{O}(\mathcal{X});\Theta) = \text{LoRA}(\text{Transformers}(\mathbf{O}(\mathcal{X});\Theta))$
        \STATE $\mathbb{Z}(n,l,k|\mathbf{O}(\mathcal{X});\Theta)$ from (\ref{optimalbeamprediction})
        \STATE $\text{LOSS}_{\text{LLM}}= \sum_{k \in \mathcal{K}} a_k \text{loss}_{n}(\mathbb{Z})$ converges
        \ENDFOR
    \RETURN Final beam selection $\hat{\mathbf{F}}$, PA positions $\hat{\mathbf{X}}$, and transmit beamforming $\hat{\mathbf{W}}$
    \end{algorithmic}
\end{algorithm}

\subsection{Computational Complexity Analysis}

The computational complexity of Algorithm~\ref{alg:llm_pass1} and Algorithm~\ref{alg:llm_pass2} is determined by the main modules: feature embedding, codebook update, LLM-based codeword selection, beamforming optimization, and PA position update.

\textbf{1) Multimodal Feature Extraction:}
The extraction of tokens from $K$ users and $N_P$ patches via the token mapping module incurs a complexity of $\mathcal{O}(N_P (s^{\text{patch}})^2)$ per iteration, where $s^{\text{patch}}$ denotes the feature dimension.

\textbf{2) LLM-Based Codebook Generation and Update:}
For each codeword $\mathbf{f}_n$, the update context involves summing over all relevant PAs and evaluating gradients concerning to the SINR. For a codebook size $N$ and $L$ PAs per waveguide, the complexity is $\mathcal{O}(N L)$ per iteration.

\textbf{3) LLM-Based Beam Selection:}
The complexity of the transformer-based LLM forward pass is $\mathcal{O}(N_P (s^{\text{patch}})^2)$, followed by a softmax in the codebook, which is $\mathcal{O}(N)$. When MoE/LoRA adapters are used, the cost of each layer increases as $\mathcal{O}(E N_P (s^{\text{patch}})^2)$, where $E$ is the number of expert adapters, typically moderate $E{=}4$. The beamforming matrix $\mathbf{W}$ is updated by iterations. For $K$ users and $M=NL$ candidate PA positions, the complexity per update is $\mathcal{O}(K^2 L N)$ due to SINR evaluation across all user and pinching beamforming codeword sets. The PA position update shares a similar cost.

\textbf{5) Overall Complexity:}
Assuming $T$ outer iterations and batch size $N_{\text{batch}}$, the overall complexity per epoch is $\mathcal{O}(TN_{\text{batch}}\left[N_P(s^{\text{patch}})^2+NL+K^2LN\right])$.
All major steps can be parallelized across codewords and users, and batched for GPU acceleration.

\section{Simulation Results}
In this section, we present numerical simulations to evaluate the performance of the proposed LLM-enabled PASS framework. Unless otherwise specified, the simulation parameters are configured as follows. First, we describe the PASS configurations in communication scenarios. Then, we provide the LLM settings. Finally, we compare the proposed LLM-enabled PASS with the conventional massive MIMO based on LLM \cite{zheng2025beamllm} and standard optimization-based PASS \cite{lv2025beam}. For simplicity, massive MIMO represents the LLM-based massive MIMO, PASS represents the standard optimization-based PASS, and LLM-PASS denotes the proposed LLM-enabled PASS. All simulations are conducted on an NVIDIA V100 GPU server. The LLM and multimodal encoder are implemented using PyTorch 2.0, YOLOv11, and GPT2.0. Training employs the AdamW optimizer with a batch size of 64, an initial learning rate of $2 \times 10^{-5}$. Model selection is based on validation set loss, and early stopping is used to prevent overfitting.
\begin{itemize}
    \item \emph{PASS Configuration}: The BS is equipped with $N=4$ waveguides, each supporting $L=16$ reconfigurable pinching PAs and associated with a dedicated RF chain. In the coverage, $K=8$ users are served simultaneously. All users are assumed to be located at ground level, with a height of zero. The user positions are uniformly distributed within a spatial region defined by $x\in[0, x^{\max}]$ and $y\in[0, y^{\max}]$, where $x^{\max} = 30$ meters and $y^{\max} = 12$ meters. Let $D$ denote the total user deployment range along the $x$-axis, so that $x^{\max} \in [0, d_0^{\rm{x}}]$. 

    All waveguides and PAs are mounted at a fixed height of $d_0 = 10$ meters. The spacing between adjacent waveguides is set to $d_0^{\mathrm{y}} = 3$ meters, and the maximum length of each waveguide is $x^{\max} = 30$ meters.
    
    For the channel and transmission settings, the maximum BS transmit power is $P_0 = 20$ dBm, and the receiver noise power is $\sigma^2 = -80$ dBm. The carrier frequency is $f_c = 15$ GHz, and the effective refractive index of the dielectric waveguide is $n_{\mathrm{eff}} = 1.4$ \cite{gan2025NOMAPASS}. The minimum required SINR for each user is set to $20$ dB. These settings are used throughout all simulations unless explicitly noted otherwise.
    \item \emph{LLM Setting}: For the proposed LLM-based PASS framework, we employ a multi-modal LLM as the core inference engine. The LLM is pretrained GPT2 on ImageNet and further fine-tuned for the downstream beam training task. The model integrates multimodal features from camera images and temporal context from historical PA position and pinching beamforming.
    The pretrained LLM input at each time step includes a high-dimensional token via a YOLO-based encoder, i.e., YOLOv11, and a corresponding beam index \cite{zheng2025beamllm,zhou2024LLM}. The hidden size and number of multi-head attention layers are set according to ablation results, with a typical configuration of 12 transformer layers and a hidden dimension of 768.
    \item \emph{Dataset}: The dataset is built upon Scenario 8 of the DeepSense 6G dataset, which is partitioned into $70\%$ for training, $10\%$ for validation, and $20\%$ for testing \cite{alkhateeb2023deepsense6G}. This scenario contains several data sequences, corresponding to individual vehicle passes. Each sequence consists of a time-aligned camera image sequence and its associated beam index and power. During data collection, the BS synchronizes camera images and power measurements for all candidate beams at each time step. These multimodal data streams are aligned to enable frame-level beam selection and training. To prepare model inputs, we segment each data sequence into overlapping windows using a fixed sliding window size of $13$. For each window, the input to the LLM encoder is $\{\mathbf{B}_k\}, k \in \mathcal{K}$ during overall time slots $T$. During training, we use an observation window of length $T$ and set the model to predict optimal beams over a horizon of $H_F$ frames. Standard experiments use $T=80$ for observation and $H_F=5$ for selection, while few-shot scenarios use $T=30$ and $H_F=10$.
\end{itemize}

\begin{figure}[htbp]
    \centering
    \includegraphics[width=3.6in]{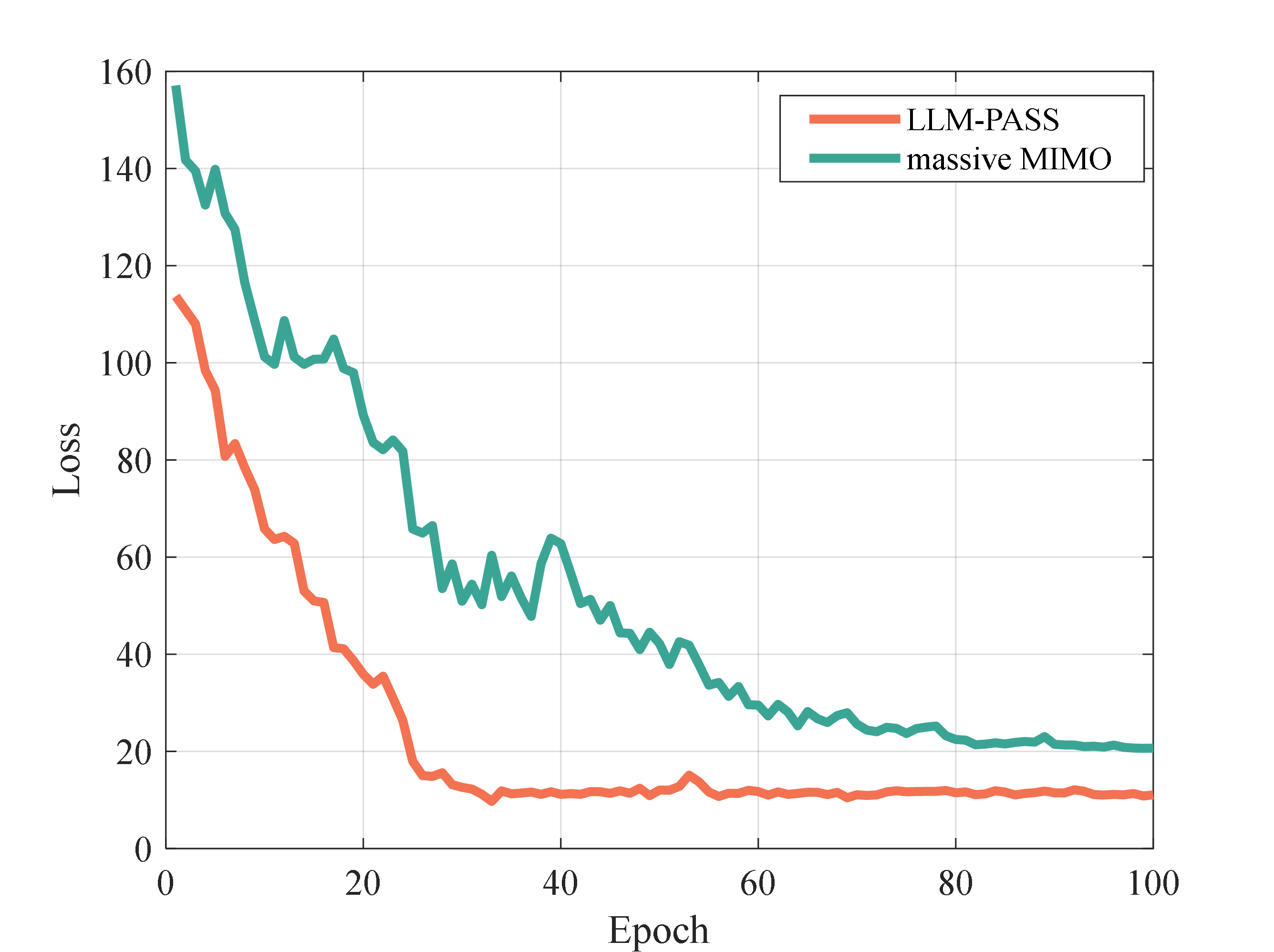}
    \caption{Comparisons of convergence behaviors of the proposed LLM-PASS and LLM-based massive MIMO.}
    \label{fig_3}
\end{figure}

Fig. \ref{fig_3} presents the convergence behaviors of the proposed LLM-PASS algorithm and the massive MIMO in terms of the training cross-entropy loss across 100 epochs. As depicted in Fig. \ref{fig_3}, the LLM-PASS exhibits a significantly faster convergence rate compared to massive MIMO. Specifically, the loss value of LLM-PASS decreases sharply within the first 30 epochs, dropping from over 110 to below 20, and subsequently stabilizes at a lower stable value, which demonstrates the superior optimization efficiency and learning capability of the proposed LLM-PASS framework. At convergence, the loss for LLM-PASS not only achieves a lower value but also shows reduced variance, implying robust generalization and stability in training. It confirms that LLM-PASS can consistently outperform massive MIMO in both convergence speed and achievable minimum loss, verifying the efficiency and practicality of the proposed solution for real-world scenarios.

\begin{figure}[htbp]
    \centering
    \includegraphics[width=3.6in]{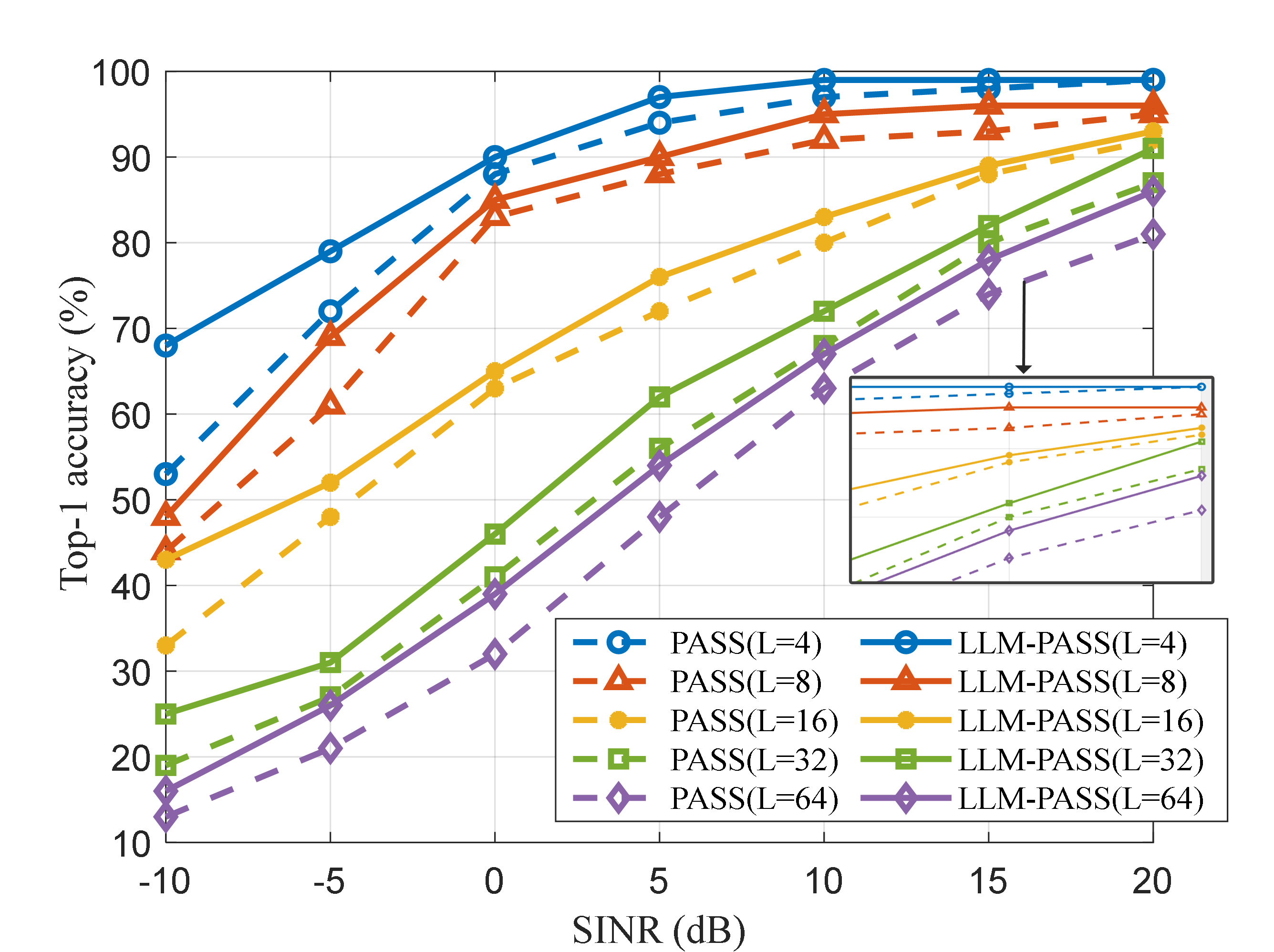}
    \caption{Comparisons of beam training Top-1 accuracy of LLM-PASS and PASS versus SINR.}
    \label{fig_4}
\end{figure}

Fig. \ref{fig_4} illustrates the relationships between Top-1 accuracy performance of both the proposed LLM-PASS and the PASS and SINRs for beam training under various numbers of pinching antennas $L=\{4,8,16,32,64\}$. It is evident from the results that, for each fixed value of $L$, the LLM-PASS consistently outperforms the PASS baseline across the entire SINR range. This demonstrates the effectiveness of leveraging LLM-driven codebook generation and beam training to improve beam selection under different channel conditions, especially poor channel conditions. At $\text{SINR}=5$ dB and $L=32$, the proposed LLM-PASS framework increases Top-1 accuracy by 10.71$\%$, compared to the PASS. In particular, the LLM-PASS method achieves Top-1 accuracy above 90$\%$ even with a low SINR with few PAs, while the conventional PASS requires significantly higher SINR or more PAs to reach similar accuracy. These results confirm that the proposed LLM-PASS is highly effective in practical scenarios, enabling robust and efficient beam training. This significant gain highlights the superiority of the LLM-driven PASS in low SINR and large-scale array.

\begin{figure}[htbp]
    \centering
    \includegraphics[width=3.6in]{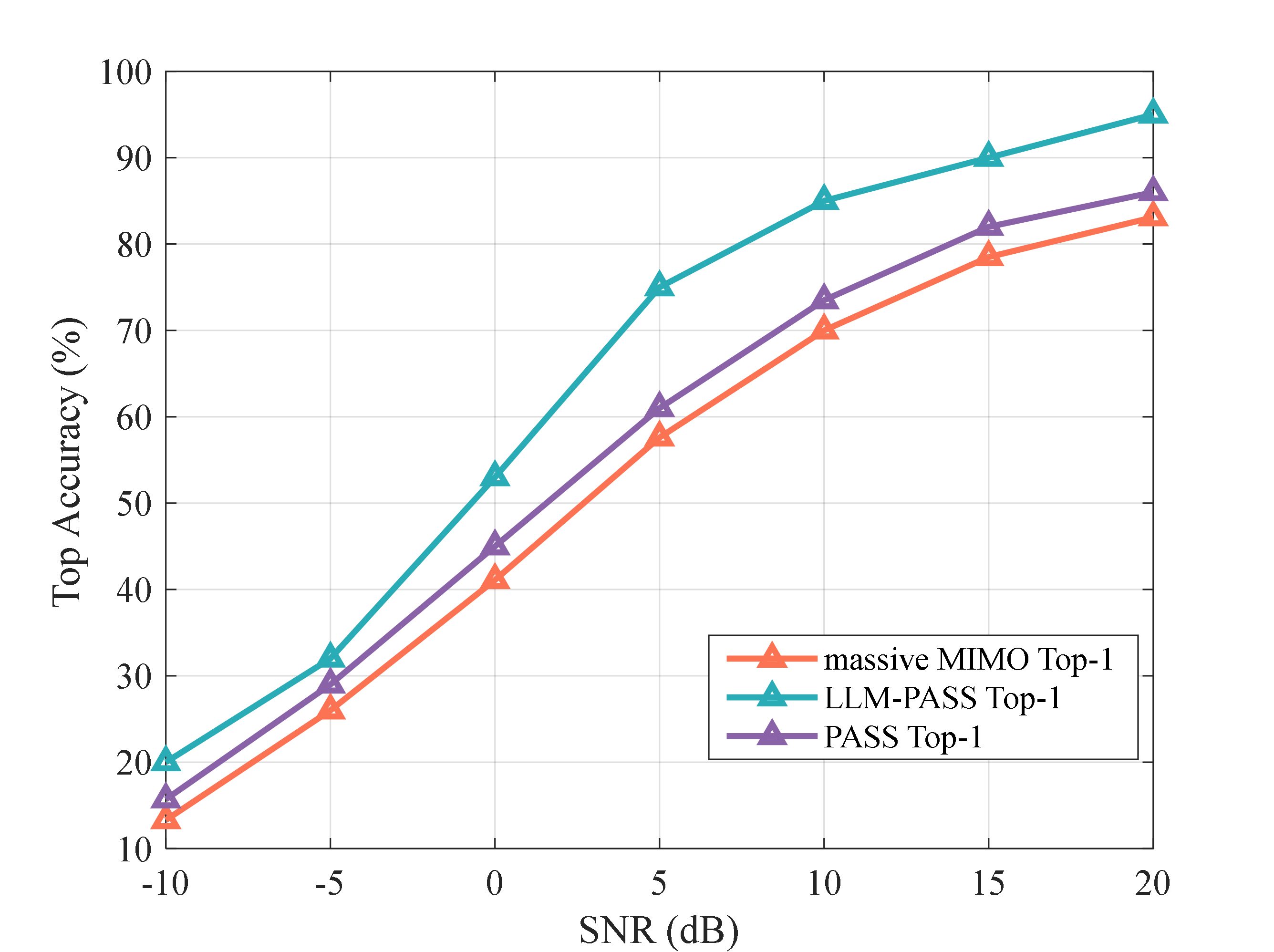}
    \caption{Comparisons of beam training Top-1 accuracy of three frameworks versus SNR.}
    \label{fig_5}
\end{figure}

Fig. \ref{fig_5} compares the Top-1 accuracy of three frameworks with different SNRs for single-user case. The proposed LLM-PASS consistently outperforms both the massive MIMO and PASS baselines in both Top-1 accuracy at all SNR levels. Specifically, LLM-PASS achieves a notably higher Top-1 accuracy at $\text{SINR}=20$ dB, LLM-PASS achieves approximately 95$\%$ Top-1 accuracy and achieves 15.16$\%$ and 10.47$\%$ gains, compared to massive MIMO and PASS, respectively. This substantial gain demonstrates the effectiveness of the LLM-PASS framework in exploiting semantic priors and spatial features to optimize beam training.

\begin{figure}[htbp]
    \centering
    \includegraphics[width=3.6in]{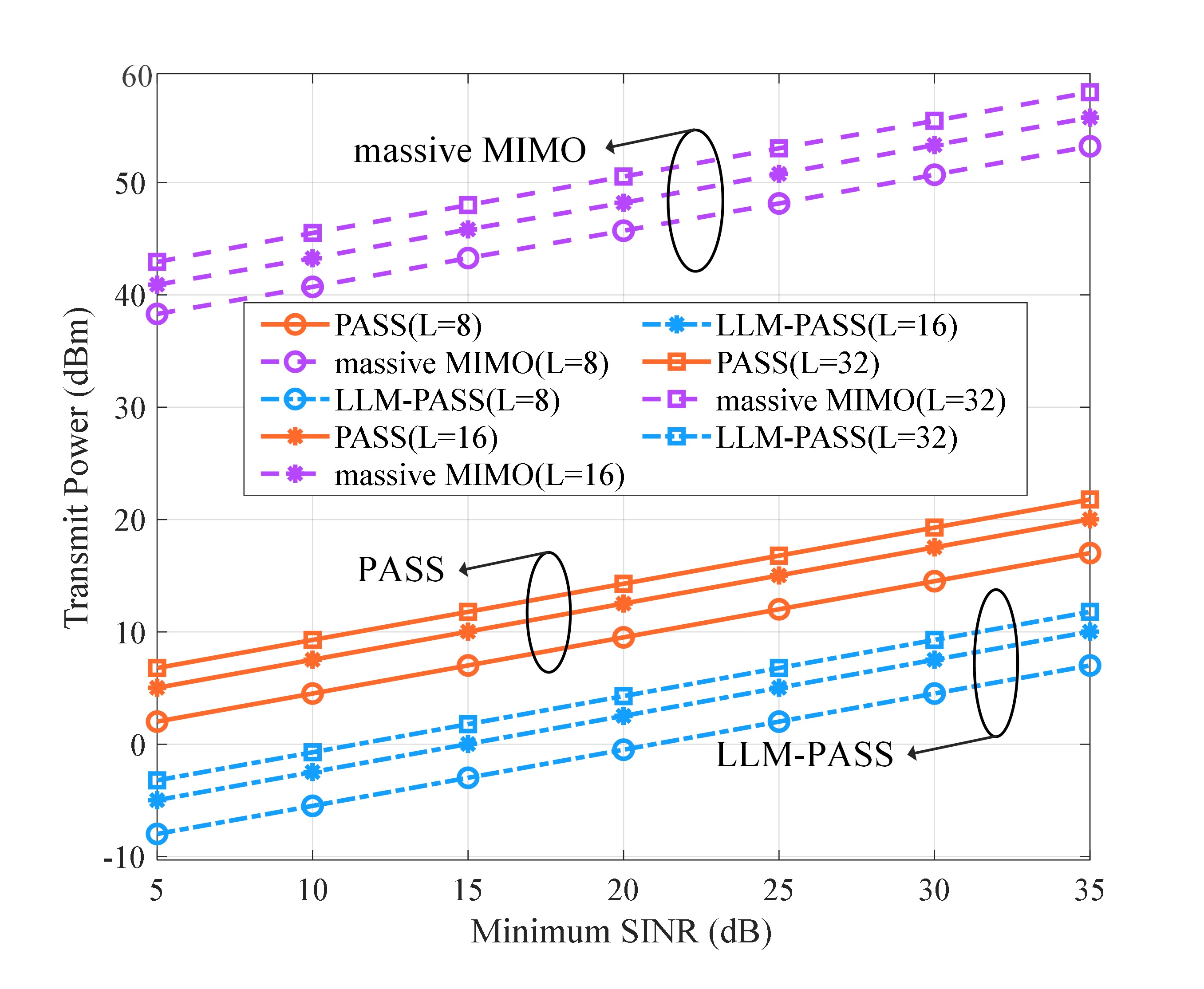}
    \caption{Comparisons of transmit power versus different minimum SINR.}
    \label{fig_6}
\end{figure}

Fig. \ref{fig_6} illustrates the performance of the proposed LLM-PASS framework and the benchmark PASS and massive MIMO in terms of the transmit power, under varying minimum SINR requirements from 5 dB to 35 dB. We can see that the transmit power increases with the higher minimum SINR requirement. As expected, the transmit power for all frameworks increases monotonically with the minimum SINR requirement, owing to the need to maintain higher signal quality at UE. At the minimum SINR of 20 dB and $L=32$, LLM-PASS reduces transmit power by 75.2$\%$ and 17.33$\%$, compared to massive MIMO and PASS, respectively. This significant reduction demonstrates the strong capability of LLM-PASS in optimizing both beam selection and antenna positioning, thereby enabling energy-efficient and robust mmWave transmission. Most importantly, these results prove that the proposed LLM-PASS not only outperforms both the conventional massive MIMO and PASS baselines in terms of transmit power, but also exhibits superior scalability as the antenna array size and SINR requirements increase, making it highly suitable for future green mmWave wireless networks.

\begin{figure}[htbp]
    \centering
    \includegraphics[width=3.6in]{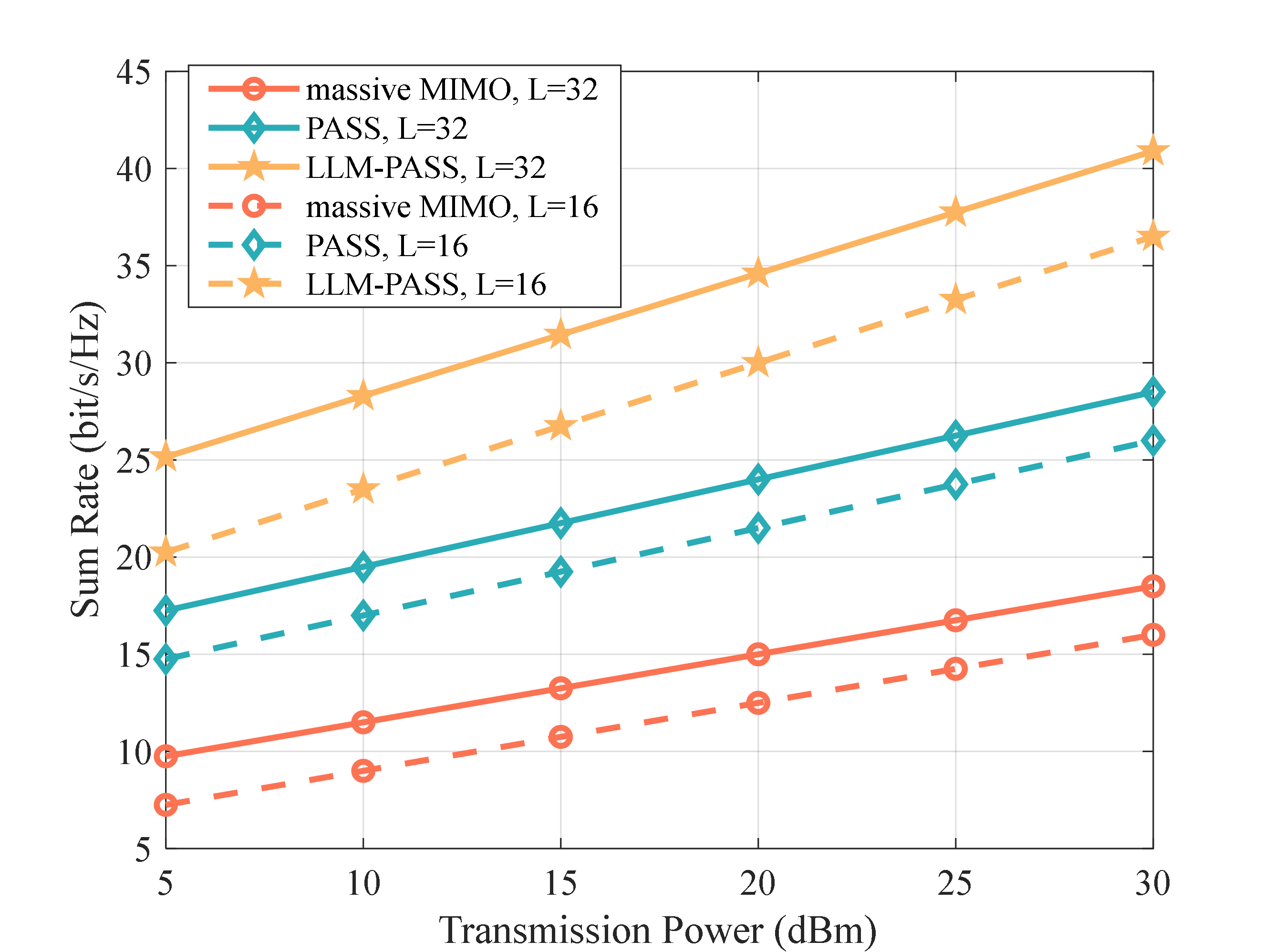}
    \caption{Comparisons of sum rate versus different transmit power.}
    \label{fig_7}
\end{figure}

Fig. \ref{fig_7} shows the achievable sum rate of three different frameworks under varying transmission power levels with two PA configurations. The achievable sum rate increases monotonically with transmission power, as expected from the fundamental Shannon capacity relationship. When transmission power is 30 dBm and $L=32$, the proposed LLM-PASS increases the achievable sum rate by 33.33$\%$ and 57.14$\%$, compared to PASS and massive MIMO, respectively. The performance gap between LLM-PASS and the baselines widens as the transmission power increases. This performance order is observed for both $L=16$ and $L=32$, confirming the superiority of the LLM-driven design in exploiting both spatial diversity and intelligent codebook optimization.

\begin{figure}[htbp]
    \centering
    \includegraphics[width=3.6in]{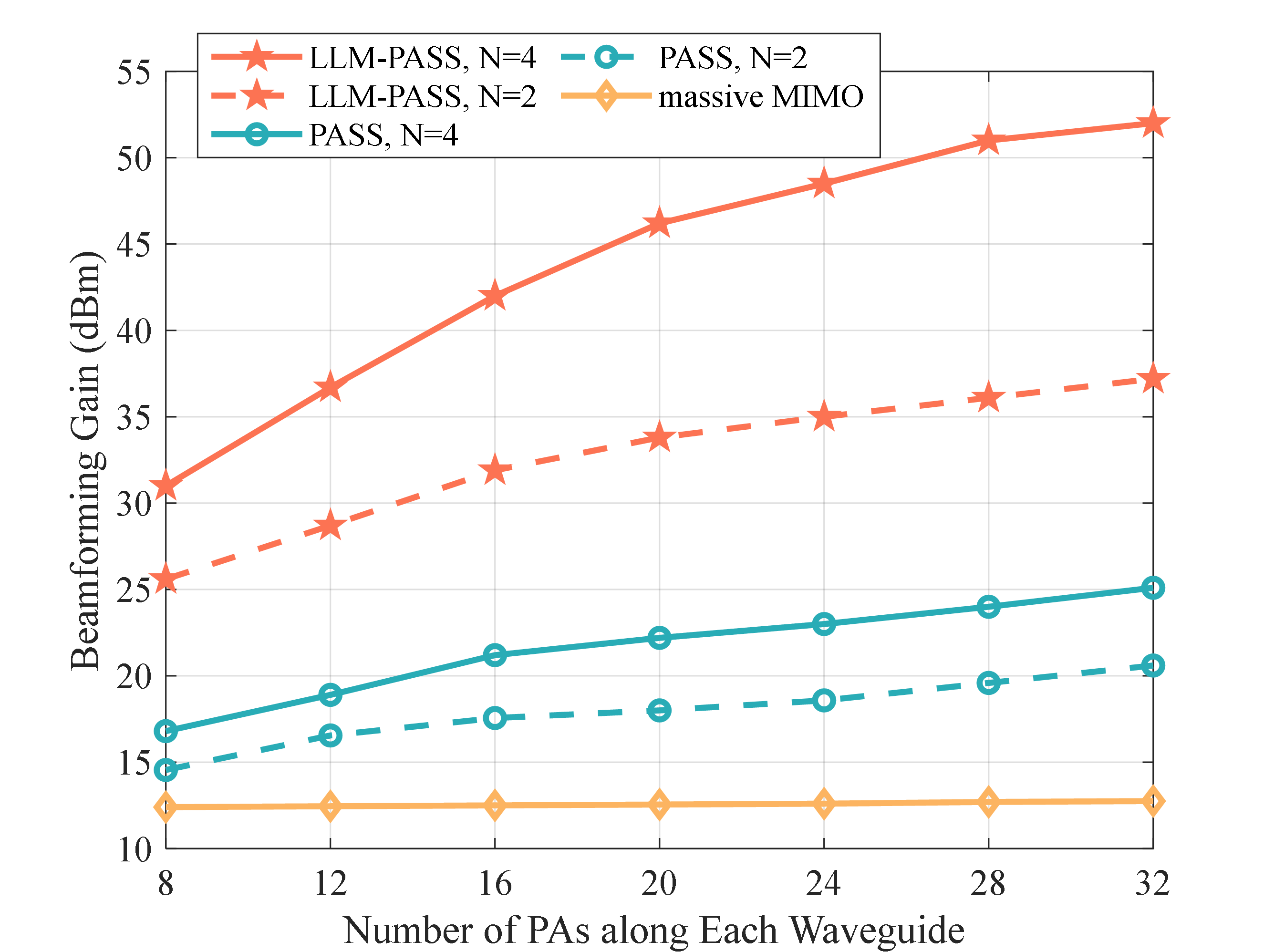}
    \caption{Comparisons of beamforming gain versus different numbers of PAs.}
    \label{fig_8}
\end{figure}

Fig. \ref{fig_8} illustrates the beamforming gain achieved by three frameworks, the proposed LLM-PASS, PASS, and massive MIMO with UPA, as the number of PAs varies. In this case, the number of antennas in massive MIMO is same as the $N\times L$. The proposed LLM-PASS approach achieves the highest beamforming gain across all configurations, with its beamforming gain increasing rapidly as the number of PAs per waveguide grows. When $N=4$ and $L=32$, LLM-PASS achieves a beamforming gain of over 50 dB, which increases this gain by 51.92$\%$ and 75.96$\%$ than PASS and the conventional massive MIMO, respectively. In stark contrast, the massive MIMO with UPA exhibits almost flat beamforming gain regardless of the number of PAs, remaining below 15 dB even for large antenna arrays. This result underlines the advantages of PASS framework in fully leveraging the hardware potential and spatial degrees of freedom provided by dense antenna deployments.

\begin{figure}[htbp]
    \centering
    \includegraphics[width=3.6in]{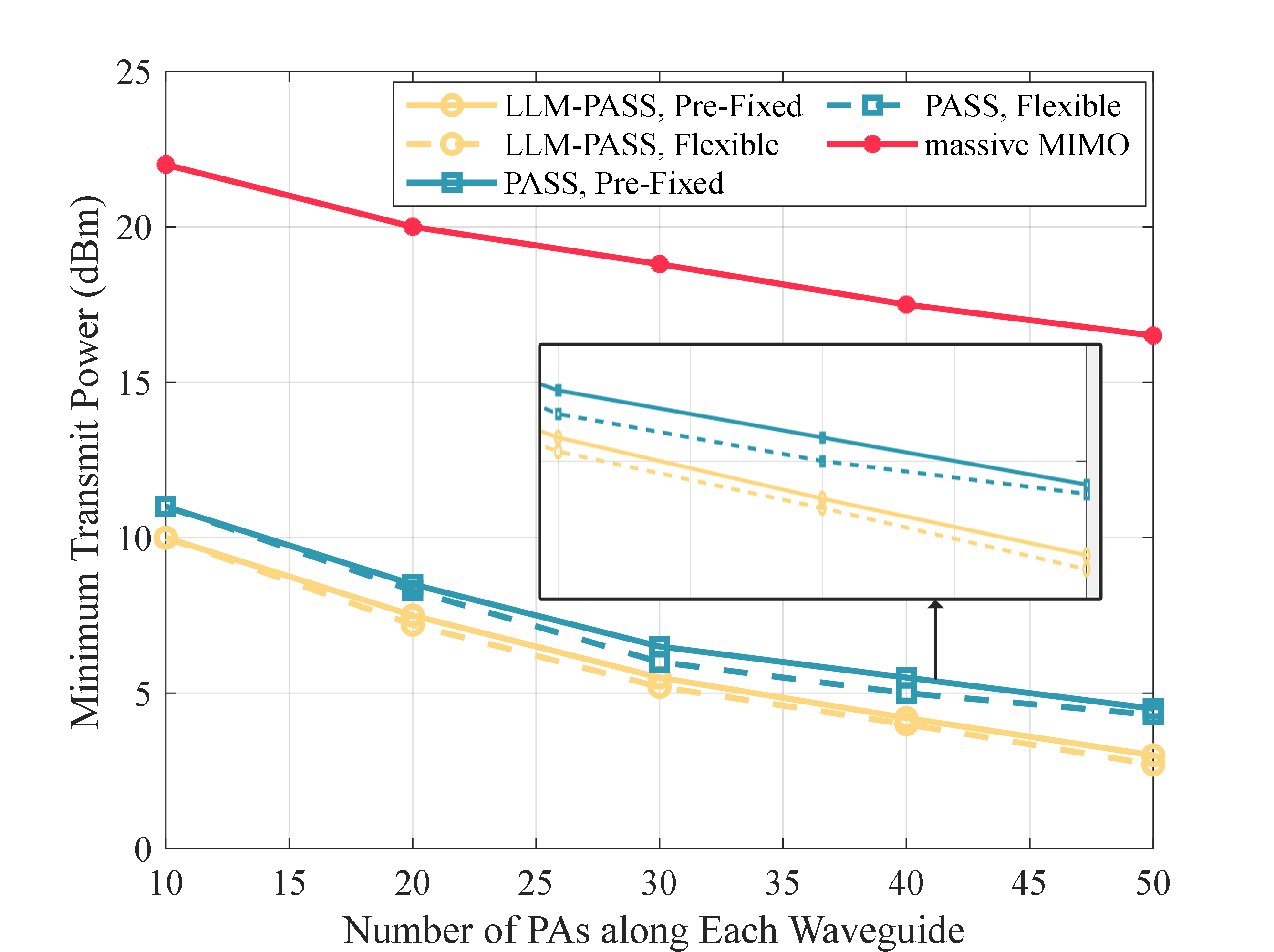}
    \caption{Comparisons of minimum transmit power versus different numbers of PAs.}
    \label{fig_9}
\end{figure}

Fig. \ref{fig_9} illustrates the relationship between the minimum required transmit power and the number of PAs under different system configurations, including LLM-PASS (Pre-Fixed and Flexible), standard PASS (Pre-Fixed and Flexible), and conventional massive MIMO. It can be observed that as the number of PAs increases, the minimum transmit power required to satisfy the communication requirements consistently decreases across all schemes. This trend highlights the significant benefits of spatial diversity and increased degrees of freedom enabled by flexible PAs, which facilitate more efficient beamforming and reduced transmit power.

\section{Conclusion}
In this paper, a novel beam training framework for PASS empowered by multimodal sensing and LLM has been proposed in downlink massive MIMO communications. The multimodal sensing data is processed by a multimodal feature encoder to produce high-dimensional tokens for efficient environmental awareness. Based on LLM, the codebook generation and beam selection mechanism has been designed to significantly reduce beam training overheads. From the obtained tokens, the pinching beamforming codebook has been generated to sample a set of pinching beamforming for sequential probing. We have considered single-user and multi-user scenarios. For single-user case, the pinching beamforming codebook generation problem has been formulated to maximize beamforming gain. Then, the optimal transmit beamforming has been obtained by MRT. For multi-user case, we have formulated a joint pinching beamforming codebook generation and beam selection problem to maximize the system sum rate with MMSE transmit beamforming strategy, which improves beamforming gains and mitigates interference. Among each user's Top-$S$ candidate beams, the optimal combination has been explored as the training labels for effective training. Then, the LLM has been trained to minimize cross-entropy loss in the supervised end-to-end training mechanism. Simulations results have demonstrated that the superiority of the proposed LLM-enabled PASS framework. Specifically, in single-user case, the proposed method has attained over 95$\%$ Top-1 accuracy in beam selection and yielded a 51.92$\%$ increase in beamforming gain compared to conventional PASS. In multi-user case, the proposed LLM-enabled PASS has resulted in up to 57.14$\%$ and 33.33$\%$ improvements in sum rate over LLM-based massive MIMO and conventional PASS.
These results have confirmed the potential of leveraging LLMs for intelligent and adaptive beam management in future reconfigurable wireless networks.

\bibliographystyle{IEEEtran}
\def\baselinestretch{1}
\bibliography{LLM_PASS}


\end{document}